\documentclass{aa}  

\usepackage{natbib,twoopt}
\usepackage{comment}
\usepackage{colortbl}
\usepackage{arydshln}
\usepackage{multirow}
\usepackage[breaklinks=true, colorlinks=true, allcolors=blue]{hyperref} 
\makeatletter
\newcommandtwoopt{\citeads}[3][][]{\href{http://adsabs.harvard.edu/abs/#3}%
{\def\hyper@linkstart##1##2{}%
\let\hyper@linkend\@empty\citealp[#1][#2]{#3}}}
\newcommandtwoopt{\citepads}[3][][]{\href{http://adsabs.harvard.edu/abs/#3}%
{\def\hyper@linkstart##1##2{}%
\let\hyper@linkend\@empty\citep[#1][#2]{#3}}}
\newcommandtwoopt{\citetads}[3][][]{\href{http://adsabs.harvard.edu/abs/#3}%
{\def\hyper@linkstart##1##2{}%
\let\hyper@linkend\@empty\citet[#1][#2]{#3}}}
\newcommandtwoopt{\citeyearads}[3][][]%
{\href{http://adsabs.harvard.edu/abs/#3}
{\def\hyper@linkstart##1##2{}%
\let\hyper@linkend\@empty\citeyear[#1][#2]{#3}}}
\makeatother
\newcommand{\xmm}{{\it XMM}}
\newcommand{\ixpe}{IXPE}
\newcommand{\nustar}{\textit{NuSTAR}}
\newcommand{\nicer}{NICER}
\def\code#1{\texttt{#1}}
\defcitealias{gianolli23}{G23} 
\usepackage{graphicx}
\usepackage{txfonts}
\usepackage{orcidlink}
\usepackage{ulem}

\titlerunning{X-ray polarization in NGC~4151}
\authorrunning{V. E. Gianolli et al.}
\begin{document} 

   \title{A second view on the X-ray polarization of NGC~4151 with IXPE}

    \author{V. E. Gianolli,$^{1,2}$\fnmsep\thanks{E-mail: vittoria.gianolli@univ-grenoble-alpes.fr}\orcidlink{0000-0002-9719-8740} S. Bianchi,$^{2}$\orcidlink{0000-0002-4622-4240} E. Kammoun,$^{2}$\orcidlink{0000-0002-0273-218X} A. Gnarini,$^{2}$\orcidlink{0000-0002-0642-1135} A. Marinucci,$^{3}$\orcidlink{0000-0002-2055-4946} F. Ursini,$^{2}$\orcidlink{0000-0001-9442-7897} M. Parra,$^{1,2}$\orcidlink{0009-0003-8610-853X} A. Tortosa,$^{4}$\orcidlink{0000-0003-3450-6483} A. De Rosa,$^{5}$\orcidlink{0000-0001-5668-6863} D. E. Kim,$^{5,6,7}$\orcidlink{0000-0001-5717-3736} F. Marin,$^{8}$\orcidlink{0000-0003-4952-0835} G. Matt,$^{2}$\orcidlink{0000-0002-2152-0916} R. Serafinelli,$^{4}$\orcidlink{0000-0002-7781-4104} P. Soffitta,$^{5}$\orcidlink{0000-0002-7781-4104} D. Tagliacozzo,$^{2}$\orcidlink{0000-0003-3745-0112} L. Di Gesu,$^{3}$\orcidlink{0000-0002-5614-5028} C. Done,$^{9,10}$\orcidlink{0000-0002-1065-7239} H. L. Marshall,$^{11}$\orcidlink{0000-0002-6492-1293} R. Middei,$^{5,12}$\orcidlink{0000-0001-9815-9092} R. Mikusincova,$^{5}$\orcidlink{0000-0001-7374-843X} P-O. Petrucci,$^{1}$\orcidlink{0000-0001-6061-3480} S. Ravi,$^{11}$\orcidlink{0000-0002-2381-4184} J. Svoboda,$^{13}$\orcidlink{0000-0003-2931-0742} F. Tombesi,$^{7,14,15}$\orcidlink{0000-0002-6562-8654}}
    \institute{Affiliations are shown at the end of the paper}

    \date{Received 24 July 2024 / Accepted 6 September 2024}

 
  \abstract
   {We report on the second observing program of the active galactic nucleus NGC~4151 with simultaneous Imaging X-ray Polarimetry Explorer (IXPE; 750 ks), \nustar\ ($\sim$ 60 ks), {\it XMM-Newton} ($\sim$75 ks), and NICER ($\sim$65 ks) pointings. 
   NGC~4151 is the first Type-1 radio-quiet Seyfert galaxy with constrained polarization properties for the X-ray corona. Despite the lower flux state in which the source was re-observed and the resulting higher contribution of the constant reflection component in the \ixpe\ energy band, our results are in agreement with the first detection. From polarimetric analysis, a polarization degree $\Pi$ = 4.7 $\pm$ 1.3\% and polarization angle $\Psi$ = 77\degr $\pm$ 8\degr east of north (68\% c.l.) were derived in the 2.0 -- 8.0 keV energy range. Combining the two observations leads to polarization properties that are more constrained than those of the individual detections, showing $\Pi$ = 4.5 $\pm$ 0.9\% and $\Psi$ = 81\degr $\pm$ 6\degr (with a detection significance of $\sim$ 4.6$\sigma$). The observed polarization angle aligns very well with the radio emission in this source, supporting, together with the significant polarization degree, a slab or wedge geometry for the X-ray corona. However, a switch in the polarization angle at low energies ($37\degr \pm 7\degr$ in the 2-3.5 keV bin) suggests the presence of another component. When it is included in the spectro-polarimetric fit, a high polarization degree disfavors an interpretation in terms of a leakage through the absorbers, instead pointing to scattering from some kind of mirror. 
   }


   \maketitle
%

\section{Introduction}
\label{sec:intro}

X-ray polarimetry has emerged as a powerful tool in unveiling the geometry of one of the innermost and most enigmatic regions of active galactic nuclei (AGN): the X-ray corona. According to the Unified Model of AGN, the corona is composed of hot \citep[with a typical temperature of kT$_e \simeq$ 10 -- 100 keV, e.g.,][]{fabian15,fabian17, tortosa18} and rarefied optically thin electron plasma situated in the innermost region of the accretion flow.
This region up-scatters the UV photons emitted by the disk, producing the primary X-ray emission \citep{sunyaev80,zdziarski00}. The polarization signal of the coronal emission is very sensitive to the geometry of the scattering material \citep[e.g.,][]{Schnittman10,tamborra18,zhang19,marinucci22,ursini22,gianolli23,ingram23,tagliacozzo23}.
Multiple studies on the physical properties of the X-ray corona \citep[i.e., the optical depth $\tau$ and the electron temperature kT$_e$,][]{shapiro76} have been performed \citep[e.g.,][and references therein]{petrucci01,matt15,ricci18,tortosa18,kamraj22,kang22,serafinelli24}. However, only with the launch of the Imaging X-ray Polarimeter Explorer \citep[IXPE;][]{weisskopf22} mission together with its synergy with simultaneous observations of the Nuclear Spectroscopic Telescope Array Mission (\nustar) and the X-ray Multi-Mirror Mission ({\it XMM-Newton}) have new insights been obtained on the geometry of this region in Type-1 AGN \citep{gianolli24,marin2024}. 

One of the X-ray brightest Seyfert galaxies in the local Universe is NGC~4151.
Intensively studied, it has shown different layers of absorption from both neutral and ionized gas \citep[e.g.,][]{Beuchert2017,gianolli23}. Archival observations of the source revealed significant spectral variability in the $1-6\,\rm keV$ range.
Furthermore, the source shows transitions from optical Type 1.5 during high-flux states, where the AGN can reach a flux level as high as $F_{\rm{0.5-10\, keV}}$ $\sim$ 2.8$\times$ 10$^{-10}$ erg s$^{-1}$ cm$^{-2}$, to optical Type 1.8 during low-flux states \citep[$F_{\rm{0.5-10 \; keV}}$ $\sim$ 8.7$\times$ 10$^{-11}$ erg s$^{-1}$ cm$^{-2}$; see][]{antonucci83,shapovalova12,Beuchert2017}.

NGC~4151 is the first Type-1 radio-quiet AGN with constrained polarization properties for the X-ray corona. From polarimetric analysis, a polarization degree (PD) of $\Pi=4.9 \pm 1.1$\% (68\% confidence level; c.l.) and a polarization angle (PA) of $\Psi=86\degr \pm 7\degr$ have been measured with a significance above 99.99\% \citep[][``G23'' hereafter;]{gianolli23}. Given that the primary X-ray emission was the dominant component in the \ixpe\ energy band and the reflection, arising from the reprocessing of the primary continuum off surrounding material (such as the torus, as favored by the analysis in G23, or the accretion disk) and which contributed only up to 6\%, the observed PD and PA refer to the primary continuum. When taken into account in the spectro-polarimetric analysis, and setting the primary continuum and reflection PAs to differ by 90 degrees, a PD = 7.7 $\pm$ 1.5\% and PA= $87\degr \pm 6\degr$ for the X-ray primary emission are found. 
Numerical simulations in \citetalias{gianolli23} show that, based on the observed X-ray polarization properties, a radially extended slab geometry is preferred for the corona. 
Interestingly, a further component with different polarization properties may be the cause of the observed change in polarization properties in the 2-3.5 keV band (PD = 4.3 $\pm$ 1.6\% and PA = $42\degr \pm 11\degr$). 
Due to its high variability in the X-ray band, NGC~4151 is an excellent candidate for studying potential changes in the observed polarization properties.

In the following, we present the spectral and spectro-polarimetric analyses of the simultaneous IXPE, \nustar, and {\it XMM-Newton} data and the results of the Neutron Star Interior Composition Explorer \citep[NICER;][]{Gendreau12,Arzoumanian14,gendreau16} monitoring. 
The structure of this paper is as follows. In Sect.~\ref{data}, we provide details on the observations and data reduction processes for \ixpe, {\it XMM-Newton}, \nustar, and \nicer. In Sect.~\ref{analysis}, we present the findings of the polarimetric, spectral, and spectro-polarimetric analyses. In Sect.~\ref{discussion} the results are discussed.

\section{Observing campaign and data reduction}\label{data}

The second \ixpe\ observation of NGC~4151 was performed during the first cycle of the \ixpe\ General Observer program.\footnote{\url{https://heasarc.gsfc.nasa.gov/docs/ixpe/proposals/ao1/c1_targets.html}} Due to the long exposure time (750 ks), it has been divided into two segments: 23-28 April and 15-26 May, 2024. We used the level two event files, which are calibrated adopting a standard \ixpe\ pipeline from the Science Operation Center (SOC).\footnote{\url{https://heasarc.gsfc.nasa.gov/docs/ixpe/analysis/IXPE-SOC-DOC-009-UserGuide-Software.pdf}}
We applied the background rejection method following \citet{dimarco23}'s guidelines for faint sources. To extract the source (background) data in the three detector units (DUs), a 68\arcsec\ circle (annulus with internal radius of 180\arcsec\ and outer of 240\arcsec) was adopted. The source extraction radius was derived iteratively to maximize the signal-to-noise ratio in the 2.0 -- 8.0 keV band, similar to the procedure described in \cite{piconcelli04}. 
Any remaining background contribution was subtracted during the polarimetric analysis. Subsequently, to estimate the polarization properties, we employed two methods: (1) creating a polarization cube \citep[\texttt{PCUBE}; using the \textsc{ixpeobssim} software version 31.0.1;][]{2022SoftX..1901194B} and (2) generating $I$, $Q$, and $U$ spectra using \textsc{xselect} \citep[from the \textsc{heasoft} package version 6.33.2;][]{2014ascl.soft08004N}, considering the weighted analysis method with the parameters \texttt{stokes=Neff} \citep{dimarco22}. 
Both methods adopt the latest available calibration file, specifically version 13 (20240101) in \textsc{ixpeobssim} and CALDB 20240125 for \textsc{xselect}. The Ancillary Response File (ARF) and Modulation Response File (MRF) for each detector unit were generated using the \texttt{ixpecalcarf} task and considering the same extraction radius used for the source region.
The results from the model-independent polarization analysis using \texttt{PCUBE} are presented in Sect.~\ref{pcube}, whereas the $I$, $Q$, and $U$ spectra are adopted in Sect.~\ref{spectropolarimetry} for the spectro-polarimetric analysis.

To constrain the physical properties of the X-ray corona and disentangle the contribution of each spectral component, we requested simultaneous \nustar\ and {\it XMM-Newton} observations. However, due to a \nustar\ star tracker blockage for the target and other time-constrained \nustar\ and \xmm\ observations during the second window of the scheduled \ixpe\ observation, both observations were divided into two pointings. 

\textit{XMM-Newton} observed NGC~4151 on May 21-22 (``X1'' hereafter) and on May 24-25 (``X2'' hereafter), 2024, for 14 ks and 25 ks (effective exposure times after filtering process) of exposure time using the EPIC-pn \citep{struder01}.
For both observations, the source and background spectra were extracted following \citetalias{gianolli23} extraction radii. A correction for the effective area was applied using the Science Analysis Software (SAS) keyword, \textsc{applyabsfluxcorr}, specifically designed to improve consistency with simultaneous \nustar\ data.

The \nustar\ \citep{harrison13} observations were carried out from May 22 to 23 (``N1'' hereafter) and from May 25 to 27 (``N2'' hereafter), 2024, adopting both coaligned X-ray telescopes equipped with Focal Plane Module A (FPMA) and B (FPMB). The \texttt{Nupipeline} task was employed alongside the latest calibration files from the database (CALDB 20221229) to generate and calibrate cleaned event files. The extraction radii for the source and background were set to be the same as those adopted by \citetalias{gianolli23}, namely, 2 \arcmin and 1.22 \arcmin. The net exposure times for FPMA and FPMB during N1 are 25 and 24 ks, while for N2 the values are 34 and 33 ks.
As found in the previous campaign, significant deviations from the pn spectrum are still present in the \nustar\ spectra below 4 keV. Hence, we considered \nustar\ data as only above 4 keV
\citepalias[see, e.g.,][]{gianolli23}.
We accounted for an energy shift of around 2000 km s$^{-1}$ at the iron line using \texttt{vashift} to address lingering calibration issues, which have also been observed in the past  \citep{gianolli23,ingram23,serafinelli23}.

Given that the coverage of \xmm\ and \nustar\ observations is limited to the second \ixpe\ observing window and due to the highly variable nature of NGC~4151, it is essential to track its absorption and flux states also during the first segment. Hence, we requested daily monitoring of the source with \textit{NICER} from April 22 to 30, 2024, for a total of $\sim$35 ks. The \nicer\ data were also collected between May 1 and 14 (other $\sim$30 ks), allowing us to check for any changes during the time frame between each \ixpe\ pointing.

The \nicer\ unfiltered data events were reprocessed using the \texttt{nicerl2} tool included in the \texttt{HEASoft} software package using the latest software and calibration files (February, 27, 2024). We applied the \texttt{nicerl3-spect} task to the clean event files and obtain the spectra. For the background, we used the \texttt{SCORPEON} model, which includes contributions from many physically motivated components. It takes into account sky-related components such as the cosmic X-ray background, local hot bubble, and galactic halo as well as non-X-ray background components such as cosmic rays, the South Atlantic Anomaly, trapped electrons, precipitating electrons, and low-energy storm-related electrons.

The extracted spectra were binned following the method of \citet{kaastra16}. All uncertainties are provided at a 68\% (1$\sigma$) c.l., unless otherwise specified, and upper and lower limits are reported at a 99\% (2.6$\sigma$) c.l. for one single parameter. Throughout our analysis, we considered a redshift of $z=0.003326$ \citep{wolfinger13} along with cosmological parameters $H_0$ = 70 km s$^{-1}$ Mpc$^{-1}$, $\Lambda_0$ = 0.73, and $\Lambda_m$ = 0.27.

\section{Data analysis}
\label{analysis}

\begin{figure*}
\begin{center}
\advance\leftskip-1.5cm
         \includegraphics[width=2.3\columnwidth]{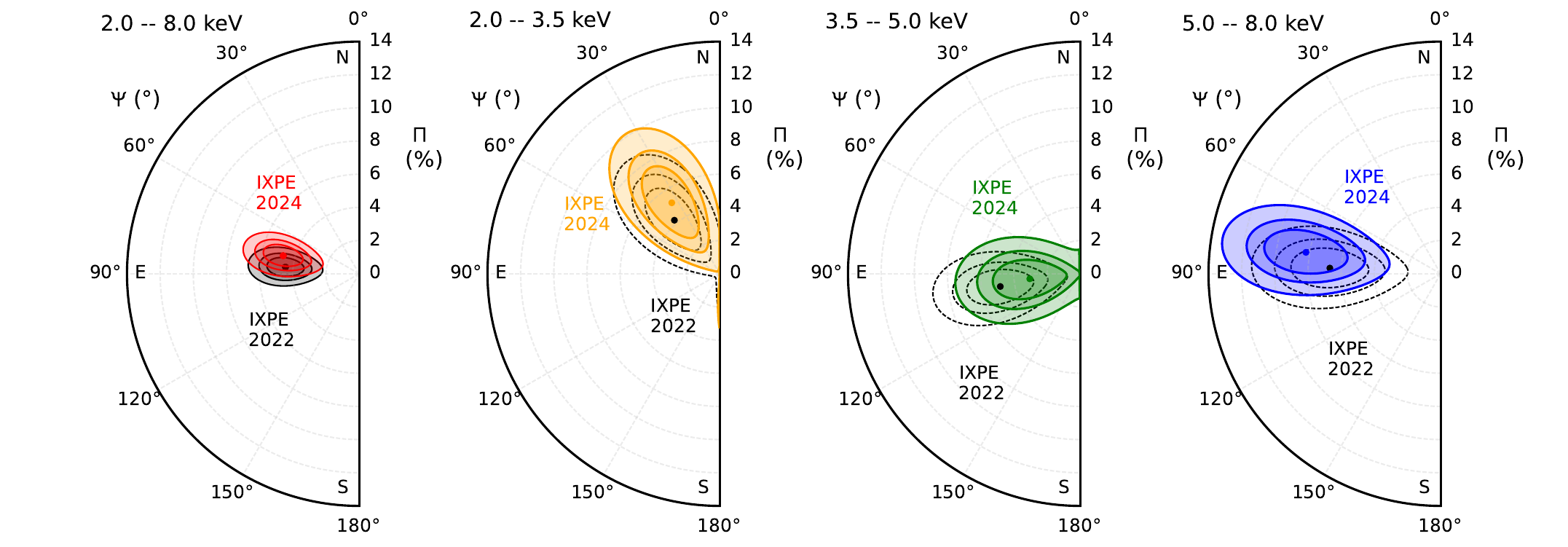}
\caption{Comparison between the first and second IXPE observations. The polarization contours are \citep[68, 90, and 99\% c.ls. ; see eq. 32 of ][]{Muleri2022} for the PD $\Pi$ and the PA $\Psi$ with respect to the north direction. First left panel: Comparison between first (in black) and second (in red) observation considering \ixpe\ full energy band.
Second to fourth panels: Comparison between first and second observation considering the division in three energy bins.
Colors refer to the 2.0 -- 3.5 keV (black, the first observation; yellow, the second observation); 3.5 -- 5.0 keV (black, the first observation; green, the second observation); and 5.0 -- 8.0 keV (black, the first observation; blue, the second observation) energy ranges, respectively. } 
\label{fig:energy_contour}
\end{center}
\end{figure*}

\subsection{IXPE polarimetric analysis}\label{pcube}

We report the results of the unweighted \texttt{PCUBE} analysis on the second \ixpe\ observation of NGC~4151. 
For the two segments separately, we obtained in the 2.0 -- 8.0 keV energy band (from the three combined DUs), after background subtraction, only an upper limit for the PD (i.e., $\Pi < 7.8$\%) in the first segment and a significant detection (with $\Pi=5.4 \pm 1.6$\% and PA $\Psi= 78\degr \pm 8\degr$) in the second. Given the consistency of the PDs in the two segments, we proceeded by summing them together, finding $\Pi= 4.7 \pm 1.3$\% and $\Psi= 77\degr\pm 8\degr$, with a detection significance at a 99.84\% c.l. ($\sim$ 3.2$\sigma$).
In Table \ref{tab:pol_energy} (middle part), we show the $\Pi$ and $\Psi$ obtained in the three energy bands adopted in \citetalias{gianolli23}: 2.0 -- 3.5, 3.5 -- 5.0, and 5.0 -- 8.0 keV. We found a significant detection only for the last energy bin (at $\sim$ 3.3$\sigma$), whereas we found marginal detections for the first and middle bins (at $\sim$ 2.6$\sigma$ and $\sim$ 1.2$\sigma$, respectively).

\begin{figure}
\centering         
\includegraphics[width=.7\columnwidth]{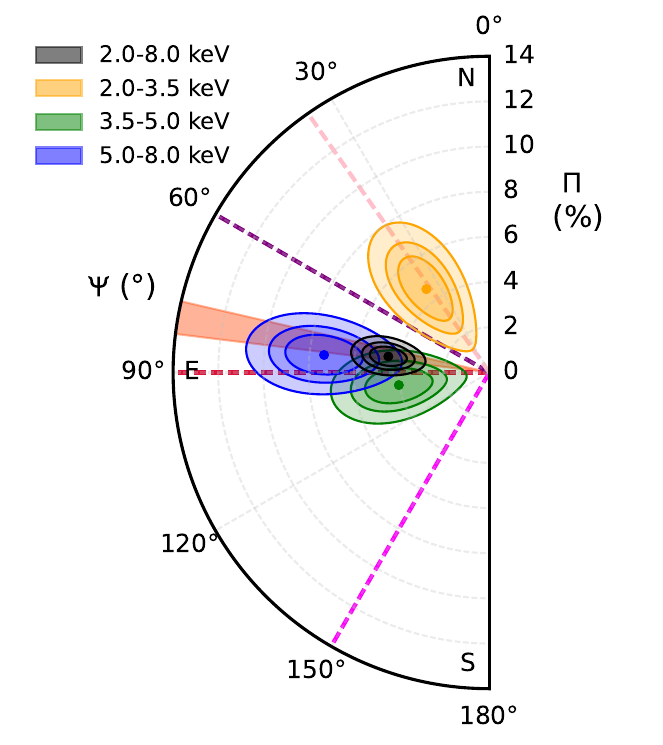}
\caption{First and second \ixpe\ observations combined. Polarization contours (68, 90, and 99\% c.ls.) for the PD $\Pi$ and the PA $\Psi$ with respect to the north direction. Colors refer to the 2.0 -- 8.0 keV (in black), 2.0 -- 3.5 keV (in yellow), 3.5 -- 5.0 keV (in green), and 5.0 -- 8.0 keV (in blue) energy ranges, respectively.
The orange region shows the direction of the radio emission. The red, purple, and pink dashed lines represent the direction of the inner few pc of the radio jet, the narrow-line region, and an unclassified feature, respectively (see Sect.~\ref{discussion}).
The magenta dashed line shows the direction of the torus, which has been identified in X-rays with a position angle of $\sim 150\degr$ \citep{Wang2011a}, appearing to coincide with the $\rm H_2$ region identified by \cite{Storchi-Bergmann2009}.
It is worth noting that the accretion disk axis could be misaligned with the torus by $\sim 20\degr$ \citep{May2020}. Furthermore, \cite{Bentz2022} describe the broad-line region as a thick disk with a $\sim 57\degr$ opening angle and a $\sim 58\degr$ inclination angle, suggesting our view skims just above the broad-line region surface.} 
\label{fig:combined_contour}
\end{figure}

To compare the results of the second observation with those of the first, we also re-extracted the \texttt{PCUBE} of the first observation, adopting the new calibration files available in \textsc{ixpeobssim}. We report the derived polarization properties in Table \ref{tab:pol_energy} (upper part).
In the left panel of Figure \ref{fig:energy_contour}, we present a comparison between the polarization contours of the first and second detections in the full energy band of \ixpe. Meanwhile, in the right panels of Figure \ref{fig:energy_contour}, we show the polarization contours obtained for the three energy bins in the first and second observations.
We observed that the polarization properties of the second observation (for both the 2.0 -- 8.0 keV energy band and the individual bins) are always well consistent with those found during the first detection. 
Hence, we combined the \texttt{PCUBE} of both observations. In Fig.~\ref{fig:combined_contour}, we report the contours of the polarimetric properties obtained in the full energy band and for the three bins. 
We observed that once the two observations are combined together, all PDs and PAs are more constrained than in the individual observations. In particular, for the full energy band of \ixpe, we obtained $\Pi= 4.5 \pm 0.9$\% and $\Psi= 81\degr\pm 6\degr$ (with detection significance $\sim$ 4.6$\sigma$). For the three energy bins, we present in Table \ref{tab:pol_energy} the PDs, which are significantly detected (at $\sim$ 3.5$\sigma$, $\sim$ 3.3$\sigma$, and $\sim$ 4.5$\sigma$, respectively) in each energy bin, and the PAs.
In the second and combined observations, the peculiar behavior detected by \citetalias{gianolli23} in the 2.0 -- 3.5 keV energy bin remains evident, and a difference of $\sim 45\degr$ between the PA in this bin and that derived for the full band is present. Meanwhile, a difference of $\sim 60\degr$ is observed between the PAs of the 2.0--3.5 keV and 3.5--5.0 keV energy bins.

\begin{table}
\centering
\caption{Polarization parameters of the first, second, and a combination of the two \ixpe\ observations for different energy bands.}
\renewcommand{\arraystretch}{1.3}
\label{tab:pol_energy}
\begin{tabular}{c c c c c}
\hline\hline
Obs &{Energy range} & {$\Pi \pm 1\sigma$} & {$\Psi \pm 1\sigma$} \\
&{[keV]} & {[$\%$]} & {[deg]} \\
\hline
1st\\
&2.0 -- 8.0 & 4.6 $\pm$ 1.2 & 85 $\pm$ 8 \\[2mm]

&2.0 -- 3.5 & 4.2 $\pm$ 1.6 & 40 $\pm$ 11\\
&3.5 -- 5.0 & 4.9 $\pm$ 1.4 & 99 $\pm$ 8\\
&5.0 -- 8.0 & 6.7 $\pm$ 2.0 & 87 $\pm$ 9\\
\arrayrulecolor{black}\cdashline{1-4}[2pt/5pt]
2nd \\
&2.0 -- 8.0 & 4.7 $\pm$ 1.3 & 77 $\pm$ 8\\[2mm]

&2.0 -- 3.5 & 5.2 $\pm$ 1.7 & 34 $\pm$ 10\\
&3.5 -- 5.0 & 3.1 $\pm$ 1.5 & 96 $\pm$ 14\\
&5.0 -- 8.0 & 8.2 $\pm$ 2.2 & 81 $\pm$ 8\\
\arrayrulecolor{black}\cdashline{1-4}[2pt/5pt]
Combined\\
&2.0 -- 8.0 & 4.5 $\pm$ 0.9 & 81 $\pm$ 6\\[2mm]

&2.0 -- 3.5 & 4.6 $\pm$ 1.1 & 37 $\pm$ 7\\
&3.5 -- 5.0 & 4.0 $\pm$ 1.0 & 98 $\pm$ 7\\
&5.0 -- 8.0 & 7.2 $\pm$ 1.4 & 84 $\pm$ 6\\
\hline
\end{tabular}
\end{table}

\subsection{Spectral Analysis: {\it XMM-Newton}, \nustar, and NICER }\label{spectral}

\begin{figure}
\centering
    \includegraphics[width=0.5\textwidth]{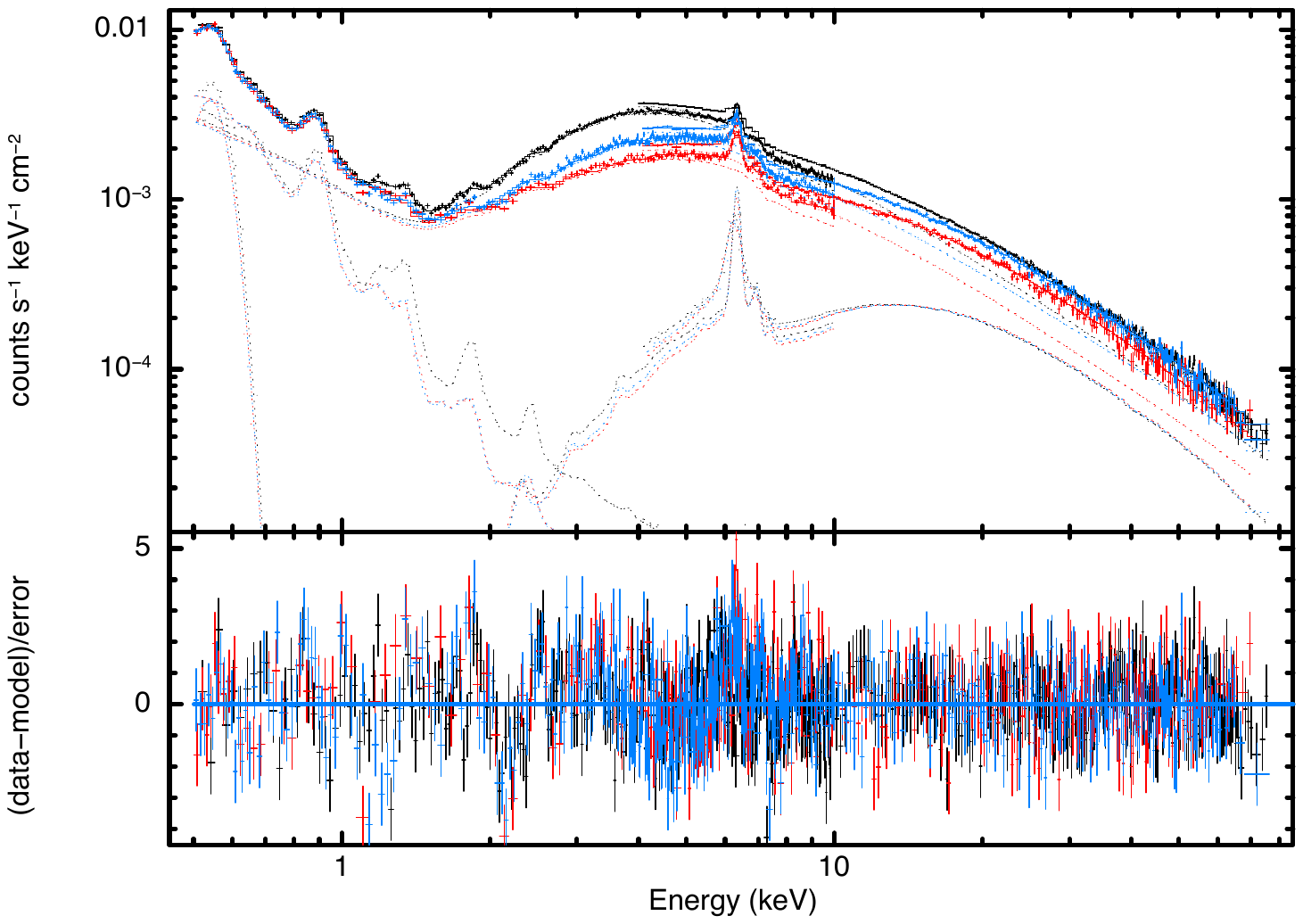}
  \caption{{\it XMM-Newton} and \nustar\ spectra with residuals. The \textit{XMM-Newton}/EPIC-pn and the grouped \nustar\ FPMA and FPMB of the first campaign are shown in black. In red, the \textit{XMM-Newton}/EPIC-pn and the grouped \nustar\ FPMA and FPMB of the second campaign (first segments) are shown. In blue, the \textit{XMM-Newton}/EPIC-pn and the grouped \nustar\ FPMA and FPMB of the second campaign (second segments) are presented. }
  \label{fig:1}
\end{figure}

To perform the spectral analysis of NGC~4151, we used \textsc{xspec} version 12.13.0 \citep{Arnaud1996}, considering the 0.5 -- 10.0 keV {\it XMM-Newton} and 4.0 -- 79.0 keV \nustar\ spectra simultaneously.
We show in Fig.~\ref{fig:1} the \xmm\ and \nustar\ spectra of the first and second observational campaigns. The spectra appear consistent below 1 keV and exhibit a similar spectral shape at high energies. Most of the spectral variability is concentrated between 2.0 and 6.0 keV, suggesting variability in obscuration.
Hence, given the complexity of modeling the X-ray spectrum of NGC~4151 \citep[e.g.,][]{weaver94,zdziarski96,yang01,derosa07,kraemer08,lubinski10,gianolli23}, we adopted the same model used in \citetalias{gianolli23}:
\code{(tbabs)*(CLOUDY + zgauss + PC*PC*WA*({BORUS} 1 + {BORUS} 2 + nthcomp))}.
We note that \texttt{tbabs} models the Galactic absorption, a thermally Comptonized continuum \citep[\texttt{nthcomp};][]{zdziarski96,zycki99} represents the primary X-ray emission, and \textsc{\texttt{BORUS}} \citep{balokovic18,balokovic19} is used for modeling the reflection component. Following the approach of  \citetalias{gianolli23}, we separated the reflection component (i.e., the emission reprocessed by a torus) from the fluorescent lines by using specific \texttt{BORUS} tables. Soft X-ray emission was accounted for by incorporating a photoionized plasma emission component generated with \texttt{CLOUDY} \citep{ferland98,Bianchi2010a}, and the residuals around the \ion{O}{vii} emission line triplet were mitigated using a Gaussian line. The different layers of absorption \citep[e.g.,][]{keck15, gianolli23} were modeled by two neutral partial-coverers (PCs; \texttt{zpcfabs}) along with a warm absorber (WA; modeled with \texttt{zxipcf} and covering a factor fixed to one).
The applied multiplicative constants (which accommodate cross-calibration uncertainties between FPM modules and EPIC-pn as well as marginal flux variability of the source, as the observations are not strictly simultaneous) are of the order of 1.20.
The source was detected in an average flux state, with $F_{\rm{0.5-10 \; keV}} = 1.07 (1.33) \times$ 10$^{-10}$ erg s$^{-1}$ cm$^{-2}$ for X1+N1 (X2+N2), which is lower than the first campaign ($F_{\rm{0.5-10 \; keV}} = 1.73  \times$ 10$^{-10}$ erg s$^{-1}$ cm$^{-2}$).

\begin{figure}
\centering
    \includegraphics[width=0.5\textwidth]{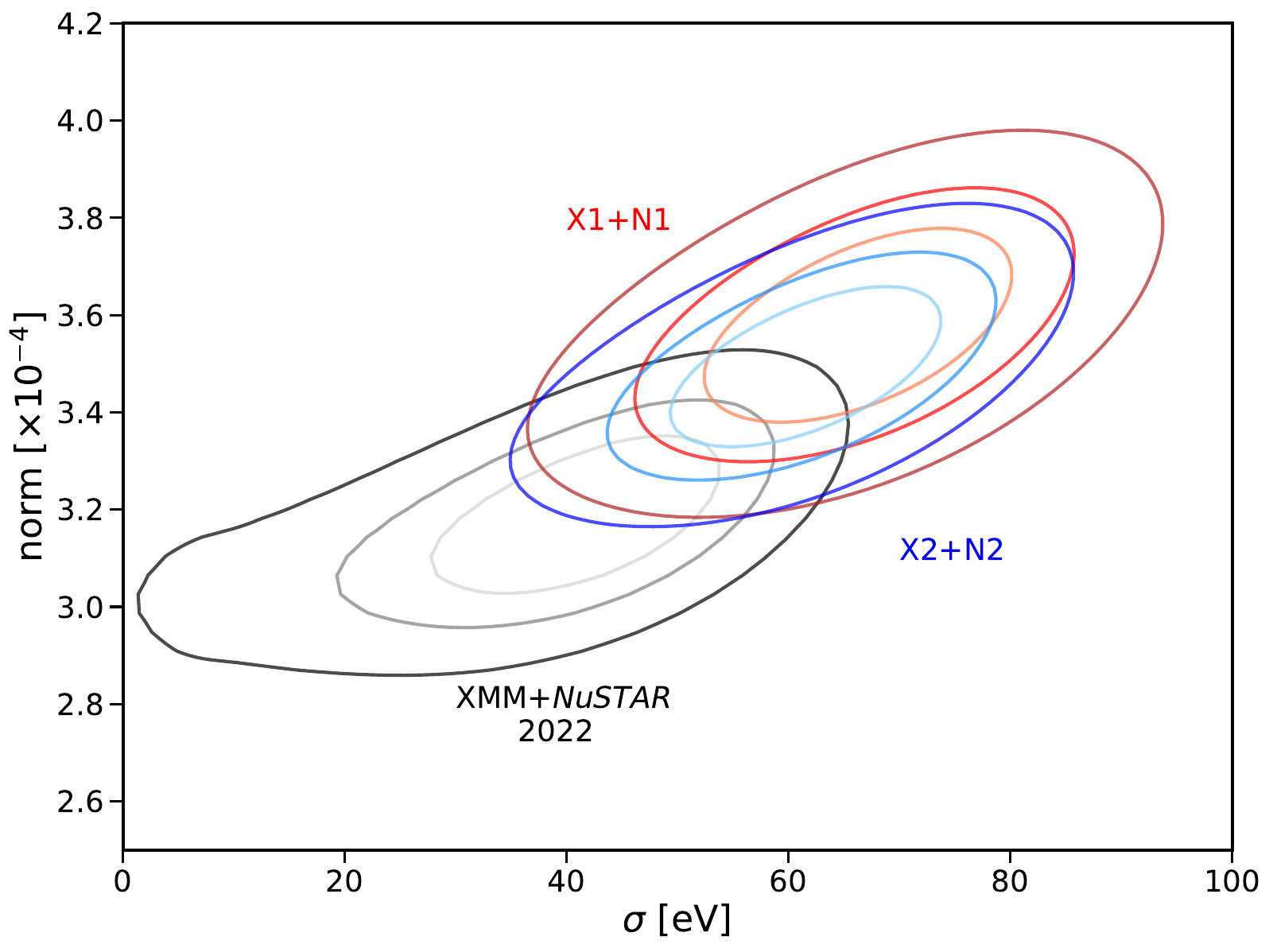}
\caption{Iron emission line: width versus normalization contour plots. Black refers to the 2022 data, red to the X1+N1 data, and blue to the X2+N2 data.}
\label{fig:contour_iron}
\end{figure}

As a first step of the spectral analysis, we analyzed the Fe K$\alpha$ emission line by focusing on the 5.0 -- 10.0 keV energy band. By doing so, we aimed at assessing the possible presence of a significant relativistic reflection component and the variability of the line. 
As found by \citetalias{gianolli23} for the first campaign, NGC~4151 does not show a clear relativistic component in the new data, which is different from what has been found in some previous observations of the source \citep[e.g.,][and references therein]{yaqoob95,zoghbi19}. The line can be modeled with a single Gaussian, but it is only slightly resolved ($\sigma\sim60-70$ eV). We find no evidence of variability in the line parameters (intensity and width), either between X1+N1 and X2+N2 or with respect to the first campaign (Fig.~\ref{fig:contour_iron}). Consequently, the equivalent width of the iron line anti-correlates with the flux of the source, being EW $=150\pm15$ ($126^{+10}_{-7}$) eV in the X1+N1 (X2+N2) data and EW $=100\pm6$ eV in the 2022 data.

Subsequently, we considered the full broadband X1+N1 and X2+N2 data, and we fit them together with the \xmm+\nustar\ data from the first campaign in order to identify any variability in the spectral components.
We assumed that the reflection component did not change, as evidenced by the absence of variation in the Fe K$\alpha$ emission line.
Meanwhile, we allowed the photon index and the normalization of the primary continuum to vary. Furthermore, all absorbers (both neutral and ionized) show a significant variation between X1+N1 and the 2022 campaign, with only the first neutral PC (column density and covering factor) differing between X1+N1 and X2+N2 (see Table \ref{tab:1}). We also observed a small but significant change for the parameters of the photoionized gas that models the soft X-ray emission, confirming a similar variability noted by \cite{zoghbi19}.
In the end, we obtained an acceptable best fit with $\chi^2$/d.o.f = 5878/5208. We show the best fit parameters in Table \ref{tab:1} (those of the 2022 campaign are also reported) and the spectra with residuals in Fig.~\ref{fig:1}. 

Whereas in \citetalias{gianolli23}, the reflection (reprocessed continuum and fluorescent lines) contribution was only about 6\% in the 2.0 -- 8.0 keV energy band, we find it to be 26\% for X1+N1 and 21\% for X2+N2. The contribution of the Fe K$\alpha$ line is 8\% (7\%) in the 5.0 -- 8.0 keV for the two new observations. 
We note that the observed trend is in agreement with our assumption on the constant reflection (which is justified via the iron line), namely as the flux increases (from X1+N1 to X2+N2 and up to the 2022 level), the contribution of the reflection decreases.
Moreover, as also obtained in \citetalias{gianolli23}, the contribution of the photoionized emission in the 2--3.5 keV band is not significant and becomes so only at lower energies. 
On the other hand, the 2--3.5 keV band manifests an excess relative to the absorbed primary continuum, which has been interpreted in this model as the primary emission leakage through the partial-coverers. Its contribution is around 15\% (22\%) in the X1+N1 (X2+N2) observation, becoming 7\% (8\%) for X1+N1 (X2+N2) in the 3.5 -- 5.0 keV and 2\% (3\%) for X1+N1 (X2+N2) in the 5.0 -- 8.0 keV bin.

\begin{table*}
\centering
\caption{Best fit from the spectral analysis.}
\label{tab:1}
\renewcommand{\arraystretch}{1.3}
\begin{tabular}{lllll}
\hline\hline
{Parameter} & {Value} & {Value} & {Value}\\
 & {1st obs} & {X1+N1} & {X2+N2}\\
\hline
\multicolumn{4}{c}{\texttt{CLOUDY \footnotesize{(Photoionized emitter)}}}\\
$\log U$ & 1.35 $\pm$ 0.01 & \multicolumn{2}{c}{1.11$^{+0.01}_{-0.02}$ $^{\textbf{+}}$} \\
$\log (N_\mathrm{H}$ / cm$^{-2}$) & 21.63 $\pm$ 0.02 & \multicolumn{2}{c}{21.54$^{+0.01}_{-0.02}$ $^{\textbf{+}}$} \\
\multicolumn{4}{c}{\texttt{PC 1 \footnotesize{(Neutral absorber 1)}}} \\
$\log (N_\mathrm{H}$ /cm$^{-2}$) & 23.02 $\pm$ 0.01 & 23.27 $\pm$ 0.01 & 23.18 $\pm$ 0.01 \\ 
C$\mathrm{f}$ & 0.78 $\pm$ 0.01 & 0.73 $\pm$ 0.01 & 0.75 $\pm$ 0.01 \\
\multicolumn{4}{c}{\texttt{PC 2 \footnotesize{(Neutral absorber 2)}}} \\
$\log (N_\mathrm{H}$ / cm$^{-2}$) & 22.64 $\pm$ 0.01 & \multicolumn{2}{c}{22.68 $\pm$ 0.01 $^{\textbf{+}}$} \\ 
C$\mathrm{f}$ & 0.95  $\pm$ 0.01 & \multicolumn{2}{c}{0.93 $\pm$ 0.01 $^{\textbf{+}}$} \\
\multicolumn{4}{c}{\texttt{WA \footnotesize{(Warm absorber)}}} \\
$\log(N_\mathrm{H}$ / cm$^{-2}$) & 23.13 $\pm$ 0.03 & \multicolumn{2}{c}{23.59$^{+0.04}_{-0.05}$ $^{\textbf{+}}$}\\ 
$\log (\xi$ / erg cm s$^{-1}$) & 4.12 $\pm$ 0.02 & \multicolumn{2}{c}{4.31 $\pm$ 0.03 $^{\textbf{+}}$} \\ 
\multicolumn{4}{c}{\texttt{\texttt{BORUS}  1/2 \footnotesize{(Neutral reflector 1/2)}}} \\
$\log (N_\mathrm{H}$ / cm$^{-2}$) & 24.45 $\pm$ 0.01 & \multicolumn{2}{c}{24.45 $^{\textbf{*}}$} \\ 
$A_{\mathrm{Fe}}$/$A_{\mathrm{Fe,\odot}}$ & 0.62 $\pm$ 0.01 & \multicolumn{2}{c}{0.62 $^{\textbf{*}}$} \\ 
norm & 0.09 $\pm$ 0.01 & \multicolumn{2}{c}{0.09 $^{\textbf{*}}$} \\ 
\multicolumn{4}{c}{\texttt{nthcomp \footnotesize{(Comptonized primary continuum)}}} \\
$\Gamma$ & 1.85 $\pm$ 0.01 & 1.88 $\pm$ 0.01 & 1.82 $\pm$ 0.01 \\
$kT_{\mathrm{e}}$ [keV] & 60 $^{+7}_{-6}$  & \multicolumn{2}{c}{60 $^{\textbf{*}}$} \\ 
norm & 0.093 $\pm$ 0.001 & 0.067 $\pm$ 0.002 & 0.071 $\pm$ 0.001  \\
\\
$\chi^2$/d.o.f. & 743/660 & 2556/2245 & 2576/2255
\\

F$_{\mathrm{2-10 \; keV}}$ / [$\times$10$^{-10}$ erg cm$^{-2}$ s$^{-1}$] & 1.680 $\pm$ 0.003 & 1.025 $\pm$ 0.004 & 1.277 $\pm$ 0.004 \\
L$_{\mathrm{2-10 \; keV}}$ / [$\times$10$^{42}$ erg s$^{-1}$] & 4.060 $\pm$ 0.006 & 2.477 $\pm$ 0.010 & 3.085 $\pm$ 0.009 \\
\hline
\end{tabular}\par
\smallskip
\footnotesize \textbf{Notes.} \\
{\bf $^*$}: During X1+N1 and X2+N2 data fitting, we fixed these values to those obtained in \citetalias{gianolli23} best fit \citepalias[we report for the first observation the uncertainties obtained in][]{gianolli23}. \\
{\bf $+$}: During data fitting, we fixed the values of X2+N2 to those obtained for X1+N1, as they do not vary when left free. \\
Normalization in units of photons s$^{-1}$ cm$^{-2}$ keV$^{-1}$.
The best fit values are adopted in the spectro-polarimetric analysis.
\end{table*}

\begin{figure*}[h!]
\centering
         \includegraphics[width=2.\columnwidth]{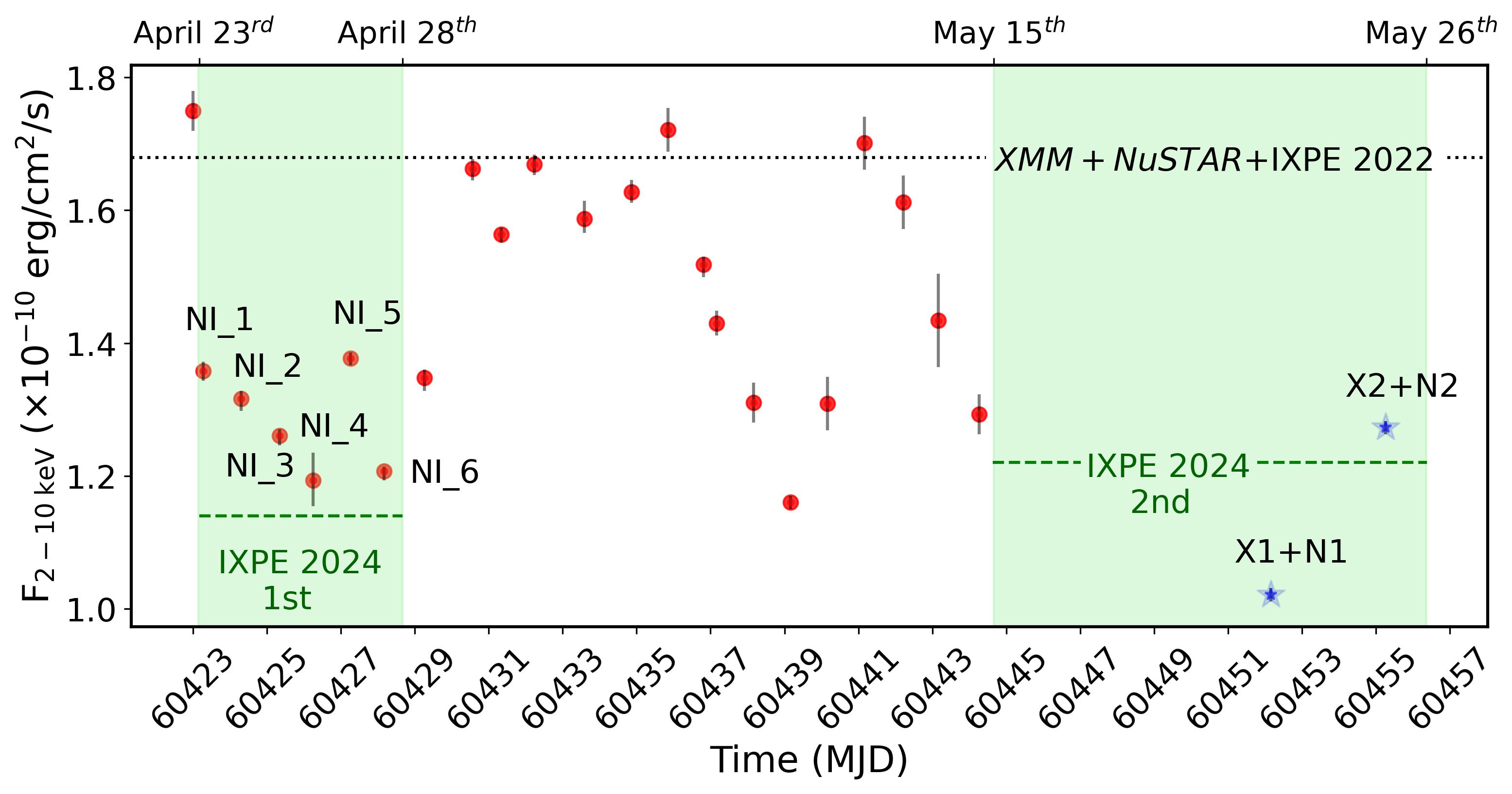}
\caption{Comparison between \nicer\ and \xmm+\nustar\ flux. The observed flux for the \nicer\ daily monitoring of NGC~4151 is shown with red dots. The light green bands show the two segments of the \ixpe\ pointing, and the dashed green lines are the respective extrapolated 2.0 -- 10.0 keV fluxes. The blue stars are the flux values obtained for X1+N1 and X2+N2.
The dotted black line represents the flux observed during the first observational campaign.}
\label{fig:nicer_flux}
\end{figure*}

\begin{figure}
\centering
         \includegraphics[width=1.\columnwidth]{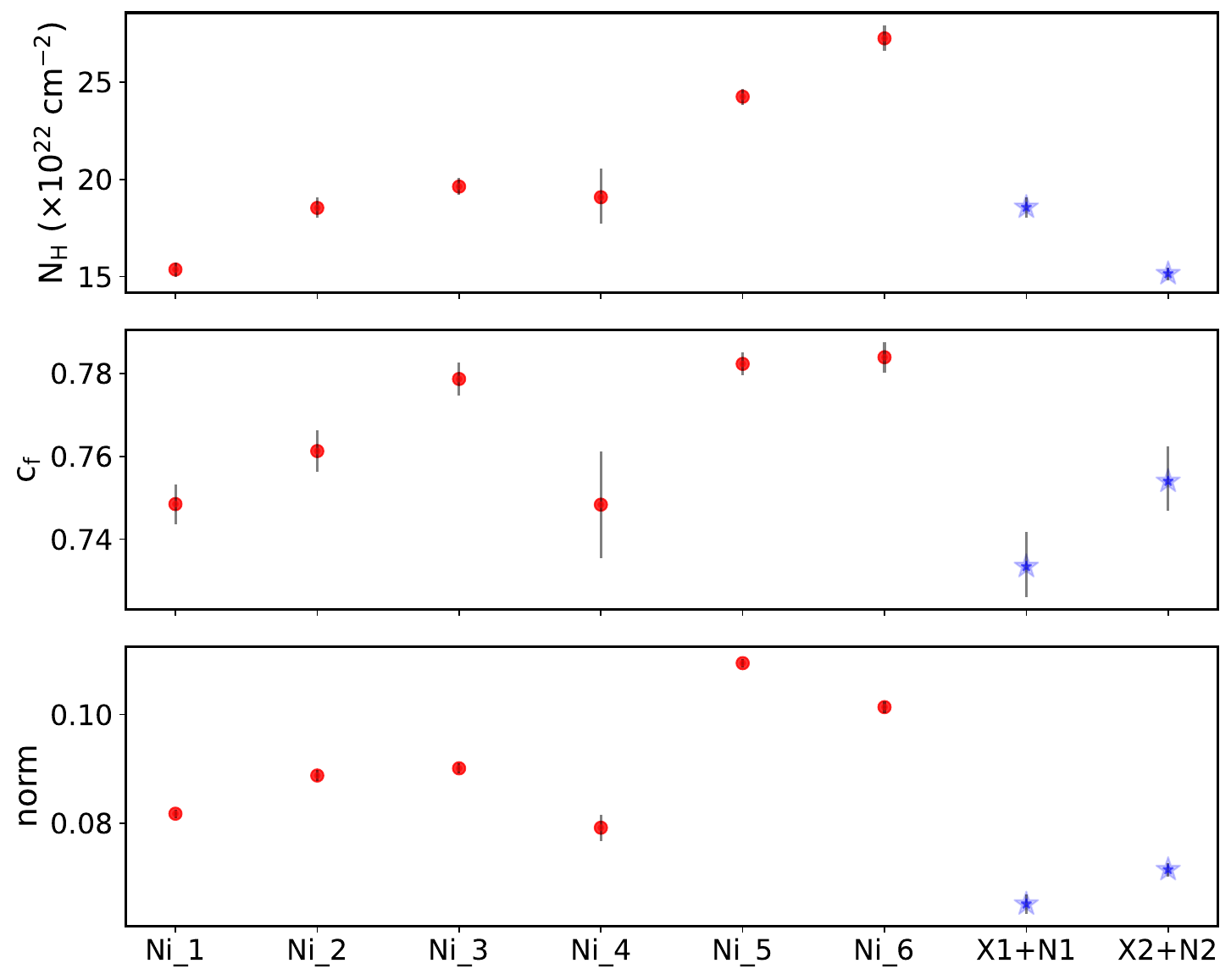}
\caption{Spectral parameter variability. We show the variations of the column density (upper panel) and covering factor (middle panel) of the first neutral partial covering along with those of the primary continuum normalization (lower panel) for the \nicer\ data (red dots) and for \xmm+\nustar\ (blue stars). }
\label{fig:variation_pc}
\end{figure}

A total of 23 \nicer\ observations were also performed continuously from the beginning of the first segment of \ixpe\ until the beginning of the second.
In Fig.~\ref{fig:nicer_flux}, we report the variation of the 2.0 -- 10.0 keV flux compared to the values observed during the past and new \xmm+\nustar+\ixpe\ campaigns.
We note that the six observations strictly simultaneous to the first segment of the new \ixpe\ observation (see light green band on the left of Fig.~\ref{fig:nicer_flux}) display a flux similar to that measured during X2+N2. 
We thus fit the six simultaneous pointings to monitor possible changes in the absorbers and continuum.
We note that due to discrepancies between the \nicer\ and \xmm\ spectra below 1 keV and because this energy band is not crucial for our purposes, we only consider data above this energy.
A good fit ($\chi^2$/d.o.f = 718/612) was obtained, allowing the normalization of the primary emission and the first neutral PC (column density and covering factor) to vary for the different spectra. In Fig.~\ref{fig:variation_pc}, we report the variation observed for the three free parameters. 

In conclusion, from the \nicer\ analysis, we confirm the presence of neutral variable absorption along the line of sight on short ($\sim$1 day) timescales during the first and between each segment of the \ixpe\ observation. The same variability, along with that of the X-ray photon index, is also evident when fitting \xmm\ and \nustar\ data. This variability extends over long timescales. Indeed, the parameters of the continuum and absorbers show variation when compared to the data taken in 2022 (see Table\,\ref{tab:1}).
However, the overall adopted model appears to remain valid. Therefore, in the spectro-polarimetric analysis discussed in the next section, we used this model for the two \ixpe\ campaigns, allowing these parameters to vary.

\subsection{Spectro-polarimetric analysis: {\it XMM-Newton}, \nustar, and IXPE}\label{spectropolarimetry}

\begin{figure}
\centering
\advance\leftskip-.5cm
    \includegraphics[width=0.5\textwidth]{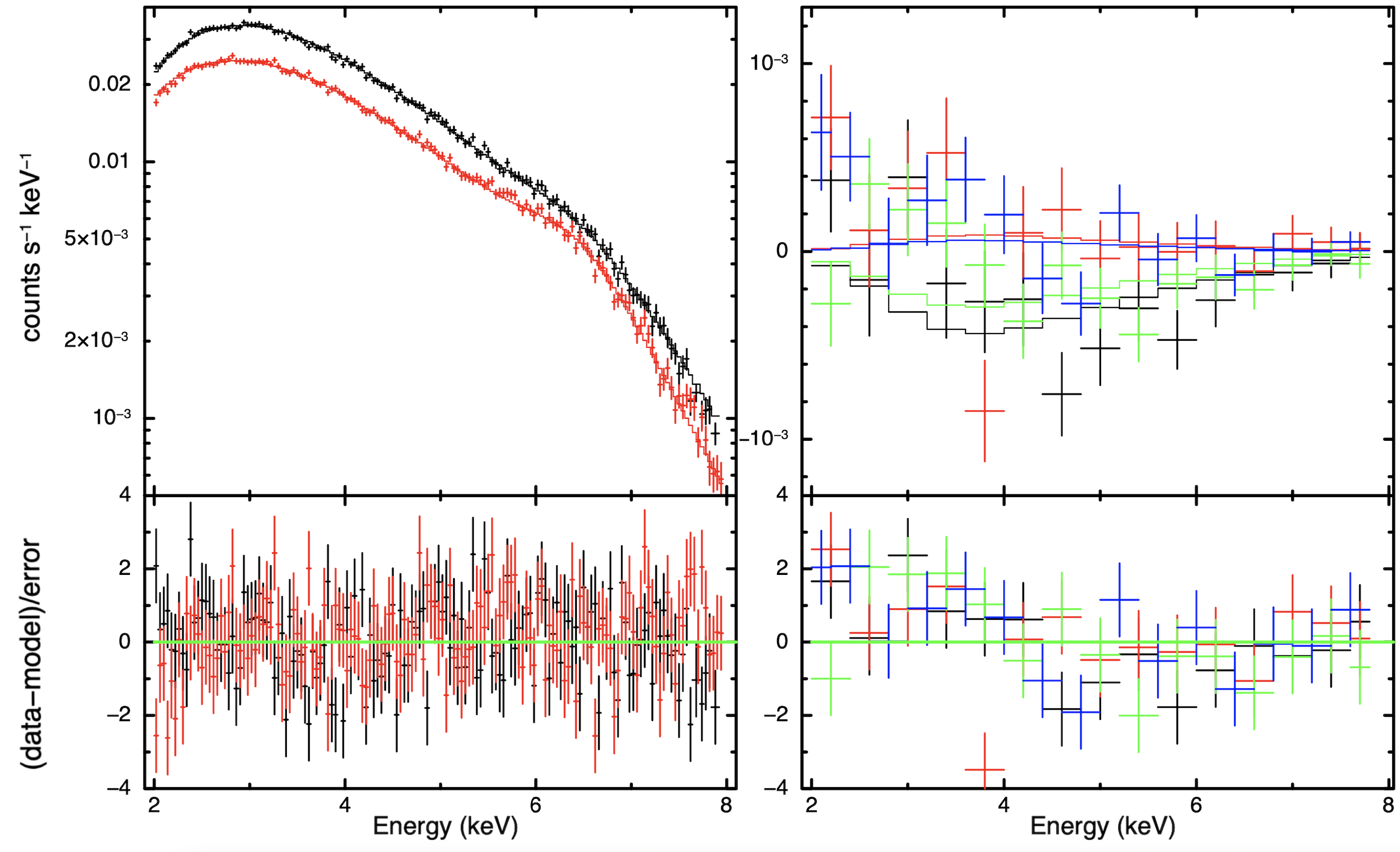}
  \caption{ IXPE spectra.
  Left panel: IXPE grouped Stokes $I$ (in black the first and in red the second observation) spectra of NGC~4151 with residuals. Right panel: $Q$ (in black the first and in green the second observation) and $U$ (in red the first and in blue the second observation) grouped Stokes spectra are shown with residuals.}
  \label{fig:ixe_spectra}
\end{figure}

The spectro-polarimetric analysis was conducted following the methodology presented in \citetalias{gianolli23}. i) We fit the IXPE data ($I$, $Q$, and $U$ spectra of the three detectors, for both observations; see Fig.~\ref{fig:ixe_spectra}) by applying the {\it XMM-Newton}+\nustar\ best fit (see Sec.~\ref{spectral}). We included an inter-calibration constant between each detector (which were free to vary and of the order of $\sim 1.00$). 
Given the variability reported in Sect.~\ref{spectral}, we left to vary the primary continuum normalization and the parameters of the first neutral partial covering for the second observation \citepalias[in comparison to those of the first observation that are set to the best fit values derived in][]{gianolli23}. 
ii) We added \code{polconst} multiplicative models in order to derive the polarization of each individual component. The PD and PA were set to zero for the \texttt{BORUS} table, accounting for the fluorescent lines and for the CLOUDY component \citepalias[see][]{gianolli23}.
We note that the \citetalias{gianolli23} spectro-polarimetric model considers a further spectral component with assigned \texttt{polconst}: the primary continuum leaking through the absorbers. 
By adopting the same configuration reported in the first paper (i.e., with the primary continuum and the reflection PAs forced to differ by 90\degr and PDs left free to vary while the PD and PA of the leaking emission were set to zero), we obtained a good fit ($\chi ^2$/ d.o.f = 1298/1222 for the 18 \ixpe\ spectra) in which the polarization is attributed to the primary continuum, with $\Pi=7.1\pm1.2$\% and $\Psi=84\degr\pm5\degr$. Only an upper limit was derived for the reflection component $\Pi<17$\% (Model 1 in Table \ref{tab:pa_pd_test}). As in \citetalias{gianolli23}, if the continuum and reflection PAs are set to be parallel (i.e., given a slab coronal geometry, the reflection is from the disk), the reflection component PD is unconstrained.

Despite the goodness of the above fit, an excess below 4 keV in the Q and U spectra was clearly visible (see Fig.~\ref{fig:ixe_spectra}, right panel). This excess drives the change in the PA in the 2.0 -- 3.5 keV energy bin (see Sect.~\ref{pcube}). 
Given the higher statistic from considering both observations, we attempted to simultaneously constrain the polarization properties of the leaking emission along with those of the X-ray continuum and of the reflection. Starting from the above spectro-polarimetric fit, we then allowed the PA and PD of the leakage to vary. Since we were unable to also constrain the polarimetric properties of the reflection component, we assumed it is either unpolarized or has PD = 20\%, with a PA parallel or perpendicular to that of the primary continuum (see Table \ref{tab:pa_pd_test}, Model 2 and Model 3). The statistical quality of the fit improved (down to $\chi^2$/ d.o.f = 1272/1221), with PD in the range $\sim10-20\%$ for both the primary continuum and the soft component in all cases and PAs of $\sim100$ and $\sim20\degr$, respectively.

\begin{table}
    \centering
    \caption{Polarization degree and angle of each spectral component for different tests conducted during the spectro-polarimetric analysis.}
    \label{tab:pa_pd_test}
    \renewcommand{\arraystretch}{1.3}
    \begin{tabular}{cccc}
    \hline\hline
        Component & PD & PA & $\chi^2$/d.o.f\\
         & {[\%]} & {[$\degr$]}\\\hline
         & \multicolumn{2}{c}{Model 1} & \\
        Primary continuum & 7.1 $\pm$ 1.2 & 84 $\pm$ 5 & \\
        Soft emission & 0 $^{fix}$ & -- & 1298/1222 \\
        Neutral reflection & $<$17 & PA$_{\rm p} \pm 90\degr$ &\\ 
        \hline
        & \multicolumn{2}{c}{Model 2} & \\
        Primary continuum & 15.6 $\pm$ 2.3 & 99 $\pm$ 4 & \\
        Soft emission & 13.1 $\pm$ 2.5 & 20 $\pm$ 6 & 1272/1221\\
        Neutral reflection & 0 $^{fix}$ & -- &\\ 
        \hline
        & \multicolumn{2}{c}{Model 3A} & \\
        Primary continuum & 19.4 $\pm$ 2.3 & 100 $\pm$ 4 & \\
        Soft emission & 14.4 $\pm$ 2.5 & 20$\pm$ 5 & 1276/1221\\
        Neutral reflection & 20 $^{fix}$ & PA$_{\rm p} \pm 90\degr$ &\\ 
        \arrayrulecolor{black}\cdashline{1-4}[2pt/5pt]
        & \multicolumn{2}{c}{Model 3B} & \\
        Primary continuum & 11.8 $\pm$ 2.5 & 98 $\pm$ 4 & \\
        Soft emission & 11.8 $\pm$ 2.5 & 20$\pm$ 6 & 1274/1221\\
        Neutral reflection & 20 $^{fix}$ & PA$_{\rm p}$ &\\
        \hline
        & \multicolumn{2}{c}{Model 4} & \\
        Primary continuum & 7.0 $\pm$ 0.9 & 93 $\pm$ 4 & \\
        Soft emission & 8.2 $\pm$ 2.2 & 22 $\pm$ 6 & 1287/1224\\
        Neutral reflection & 0 $^{fix}$ & -- &\\
        \hline
    \end{tabular}\par
\smallskip
\footnotesize \textbf{Notes.} 
PA$_{\rm p}$ is the PA of the primary continuum.
\end{table} 

\begin{figure}
\centering
    \includegraphics[width=0.4\textwidth]{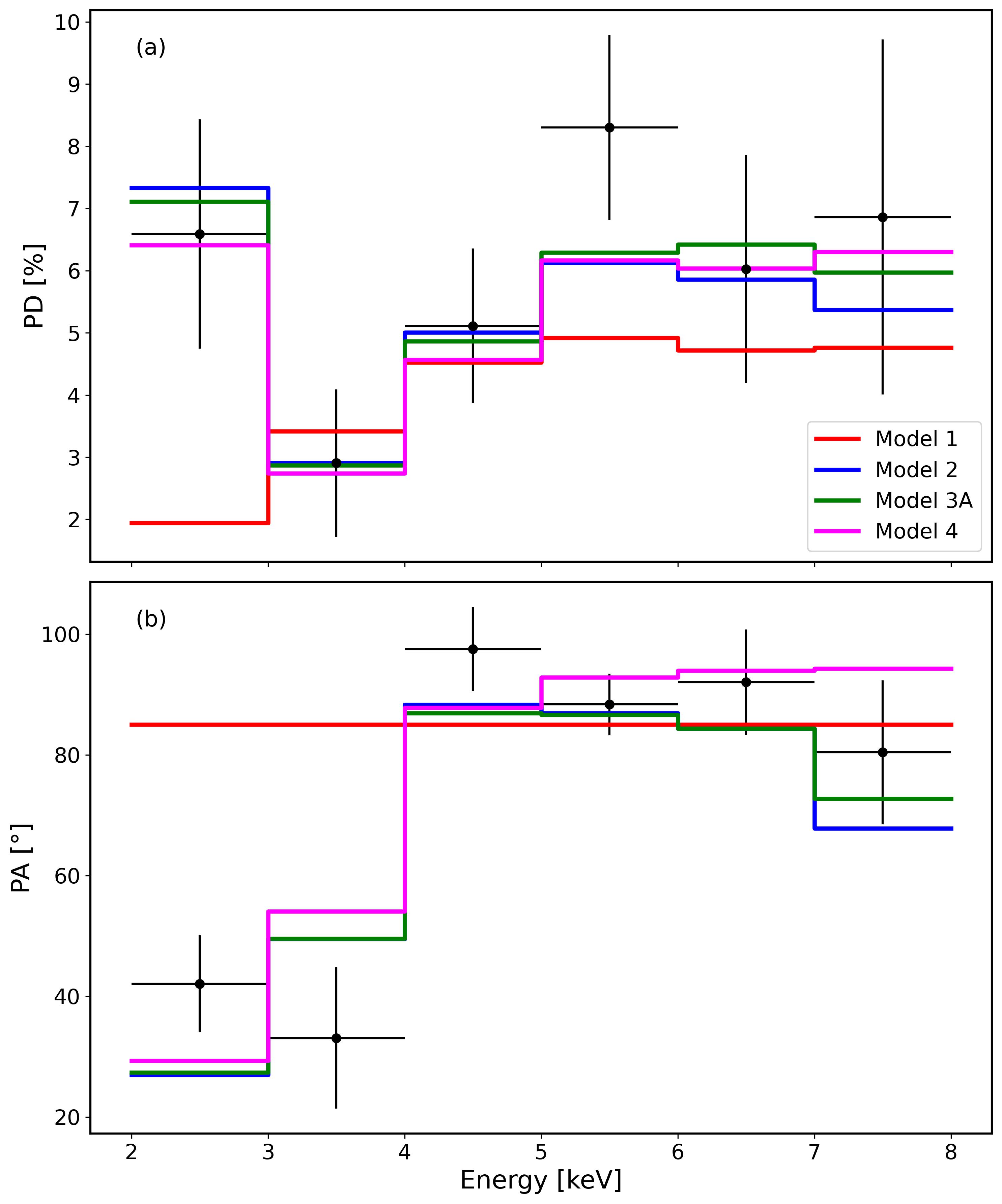}
  \caption{Polarization properties as a function of energy.
   Panel a: Polarization degree. Panel b: Polarization angle. Referring to the models presented in Table \ref{tab:pa_pd_test}, we show Model 1 in red, Model 2 in blue, Model 3A in green, and Model 4 in magenta. The \ixpe\ spectra are rebinned at an interval of 1 keV. }
  \label{fig:ixpe_spectra_rebin}
\end{figure}

We show in Fig.~\ref{fig:ixpe_spectra_rebin}, the PD and PA as a function of energy, as predicted by the models described above and compared to the observed values, rebinned at 1 keV and grouped (all detectors for both observations) for a better visualization.
Although these fits account for the switch of the PA in the soft X-rays, their physical interpretation is problematic. A PD of around 14\% for the soft component may suggest that what we observed is not a leakage through the absorbing layers but rather the reflection of the primary emission on the inner accretion flow or on a mirror close to the disk. Concerning the primary emission, a Comptonizing X-ray corona cannot produce a PD in the 16-20\% range. For instance, a slab corona, which produces the highest PDs, can reach up to PD $\sim$ 12\%, depending on the inclination \citep[see][]{ursini22}.

In order to disentangle the coronal emission from the soft component, we tested a simplified phenomenological model composed only of a power law and a blackbody (with temperature $\sim$ 0.5 keV) and with each obscured by a different layer of neutral matter (of the order of 10$^{23}$ and 10$^{22}$ cm$^{-2}$, respectively).  The two components have separate polarization properties, while the Fe K$\alpha$ line is considered unpolarized. 
From a statistical point of view, the fit is similar to the previous ones ($\chi^2$/d.o.f = 1287/1224). While still reproducing the angle switch at low energies, this model recovers a PD of $\sim7\%$ for the primary emission, with a similar PD ($\sim8\%$) for the soft component (Table \ref{tab:pa_pd_test}, Model 4, and Fig.~\ref{fig:ixpe_spectra_rebin}).

\section{Discussion and conclusions}
\label{discussion}

Our spectral analysis of the available {\it XMM-Newton}, \nustar, and \nicer\ data of NGC~4151 (see Sect.~\ref{spectral}) revealed variations in the absorbers and in the primary continuum compared to the 2022 data and between the two new sets of observations. Specifically, we observed variability in the X-ray photon index, continuum normalization, and both neutral and ionized absorbers along the line of sight on short ($\sim$ 1 day) and long ($\sim$ 1.5 year) timescales, suggesting ongoing dynamic processes in the source environment (see Fig.~\ref{fig:nicer_flux} and \ref{fig:variation_pc}, and Table \ref{tab:1}). Meanwhile, all the physical parameters of the reflection component are compatible with being constant with respect to the past data (see Table \ref{tab:1} and Fig.~\ref{fig:contour_iron}), so its contribution to the \ixpe\ energy band is higher in the new observations (see Table \ref{tab:1}). 

Despite the lower flux state of NGC~4151 and the higher contribution of the reflection component, we find no evidence for variations in the observed polarization properties, with the PD and PA being in good agreement between the first and second observations in all considered energy ranges (see the polarimetric analysis in Sect.~\ref{pcube}). This allowed for an analysis of the combined observations, which yielded better constraints for PD and PA. In particular, in the full \ixpe\ energy band (i.e., 2.0 -- 8.0 keV), we obtained $\Pi = 4.5 \pm 0.9$\% and $\Psi = 81\degr \pm 6\degr$. The latter aligns very well with the radio emission (see Fig.~\ref{fig:combined_contour}), which shows an average position angle of $\sim 77-83\degr$ \citep[e.g.,][see Fig.~\ref{fig:combined_contour}]{Harrison1986,ulvestad98,Mundell2003, Williams2017}.
Furthermore, we confirm with a higher confidence (detected at $\sim$ 3.5$\sigma$ in the combined data versus $\sim$ 2.2$\sigma$ in the individual observations) the change in PA between the low energy bin (2.0 -- 3.5 keV) and the higher bins (i.e., 3.5 -- 5.0 keV and 5.0 -- 8.0 keV) or the full band, already observed in \citetalias{gianolli23} (see Table \ref{tab:pol_energy} and Fig.~\ref{fig:energy_contour} and \ref{fig:combined_contour}). Interestingly, the PA in this soft X-ray bin (i.e., $37\degr \pm 7\degr$) appears to align with an unclassified feature identified by \cite{Draper1992} that presents an enhanced extra-nuclear polarization in the $I$-band, located at $\sim 17''$ from the center at position angles of $35\degr$ and $215\degr$ (see Fig.~\ref{fig:combined_contour}).

A spectro-polarimetric fit on both \ixpe\ observations gave results in full agreement with what is presented in \citetalias{gianolli23}, albeit with more constrained PD and PA values (i.e., $\Pi=7.1 \pm 1.2$\% and $\Psi=84\degr \pm 5\degr$) for the X-ray corona emission. As detailed in \citetalias{gianolli23}, these polarization properties are consistent with a slab or wedge geometry for the X-ray corona, as strongly reinforced by the alignment with the double-sided radio jet mentioned above.  

However, in this spectro-polarimetric fit, an excess of the Q and U data with respect to the model is apparent below 4 keV (see Fig.\ref{fig:ixe_spectra}), driving the clear change of PA at lower energies (see Fig.~\ref{fig:combined_contour} and \ref{fig:ixpe_spectra_rebin}). This observational evidence suggests the presence of an additional soft X-ray component, which could be dominated by the leakage of the primary emission through the absorbers, as modeled in our spectroscopic best fit. 
We allowed the PA and PD of the leakage to vary, assuming that the reflection component is either unpolarized or has a PD of 20\% with a PA parallel or perpendicular to the primary continuum (see Sect.~\ref{spectropolarimetry}). This improved the statistical quality of the spectro-polarimetric fit, notably modeling the PA switch at low energies (Table~\ref{tab:pa_pd_test} and Fig.~\ref{fig:ixpe_spectra_rebin}).
The PD found for the soft component is fairly high (in the range of 12-14\%), disfavoring the leakage hypothesis and rather suggesting scattering of the primary emission off the inner accretion flow or another mirror. In the latter case, this may suggest that the contribution of the NLR emission to the \ixpe\ energy band is much larger than what is predicted by our best fit model as a result of an over-simplification of the very complex spectrum of NGC~4151 around 2 keV. However, the PA attributed to this component in these spectro-polarimetric fits is around $20\degr$, which is very distinct compared to the observed position angle of $\sim 60\degr$ of the extended, biconical, NLR observed in near-IR, [\ion{O}{III}] and X-rays \citep[see, e.g.,][and references therein]{Draper1992, Wang2011a, Marin2020}.

These fits yielded a higher PA for the primary continuum, around $100\degr$, which somewhat departs from the position angle of the radio emission, although sub-arcsecond resolution observations have shown that the inner $\sim 150$pc of the radio jet is aligned with a PA of $\sim 90\degr$ \citep{Mundell2003}.
Moreover, as for the soft X-ray component, the derived PD for the primary continuum is very high (10-20\%), challenging the interpretation of this component as the intrinsic coronal emission.
Both issues are solved by completely disentangling the primary continuum from the soft component thanks to a simplified phenomenological model, composed only by a power-law and a blackbody.  This suggests again that a significant part of the spectral complexity around 2-3 keV is not due to leakage of the primary continuum but rather to an independent component.

We did not detect clear evidence of reflection off the accretion disk in the current data (see Sect.~\ref{spectral}) nor in the 2022 spectra (see G23). However, we cannot rule out the possibility that a significant part of the X-ray polarization above $\sim 4\,\rm keV$ is produced by disk reflection \citep[see, e.g.,][]{Podgorny2022, Podgorny2023} without clear spectroscopic signatures. This scenario will be explored with a self-consistent spectro-polarimetric analysis in a forthcoming publication (Dov\v{c}iak et al., in preparation).
We note here that observing NGC~4151 with \ixpe\ in a much lower flux state could break this degeneracy and provide a more comprehensive understanding of the polarimetric properties of the AGN.

\begin{acknowledgements}
      The Imaging X-ray Polarimetry Explorer (IXPE) is a joint US and Italian mission. The US contribution is supported by the National Aeronautics and Space Administration (NASA) and led and managed by its Marshall Space Flight Center (MSFC), with industry partner Ball Aerospace (contract NNM15AA18C). The Italian contribution is supported by the Italian Space Agency (Agenzia Spaziale Italiana, ASI) through contract ASI-OHBI-2017-12-I.0, agreements ASI-INAF-2017-12-H0 and ASI-INFN-2017.13-H0, and its Space Science Data Center (SSDC) with agreements ASI- INAF-2022-14-HH.0 and ASI-INFN 2021-43-HH.0, and by the Istituto Nazionale di Astrofisica (INAF) and the Istituto Nazionale di Fisica Nucleare (INFN) in Italy. This research used data products provided by the IXPE Team (MSFC, SSDC, INAF, and INFN) and distributed with additional software tools by the High-Energy Astrophysics Science Archive Research Center (HEASARC), at NASA Goddard Space Flight Center (GSFC). AT acknowledges financial support from the Bando Ricerca Fondamentale INAF 2022 Large Grant ‘Toward an holistic view of the Titans: multi-band observations of z>6 QSOs powered by greedy supermassive black holes’. RS acknowledges funding from the INAF-PRIN grant "A Systematic Study of the largest reservoir of baryons and metals in the Universe: the circumgalactic medium of galaxies” (No. 1.05.01.85.10). We thank the IXPE, \nustar, {\it XMM-Newton}, and NICER SOCs for granting and performing the respective observations of the source.
\end{acknowledgements}

\section*{Data availability}

The \ixpe\ data used in this paper are publicly available in the HEASARC database (\url{https://heasarc.gsfc.nasa.gov/docs/ixpe/archive/}). The {\it XMM-Newton} and \nustar\ data are subject to an embargo of 12 months from the date of the observations. Once the embargo expires (May 31, 2025) the data will be publicly available from the {\it XMM-Newton} (\url{http://nxsa.esac.esa.int/}) and the {\it NuSTAR} (\url{https://heasarc.gsfc.nasa.gov/docs/ nustar/nustar_archive.html}) archives. The \nicer\ data are publicly available in the HEASARC database (\url{https://heasarc.gsfc.nasa.gov/docs/nicer/nicer_archive.html}).

%
   \bibliographystyle{aa} 
   \bibliography{biblio} 

\begin{thebibliography}{1}
\expandafter\ifx\csname natexlab\endcsname\relax\def\natexlab#1{#1}\fi

\bibitem[{{Bianchi} {et~al.}(2006){Bianchi}, {Guainazzi}, \&
  {Chiaberge}}]{bianchi06}
{Bianchi}, S., {Guainazzi}, M., \& {Chiaberge}, M. 2006, \aap, 448, 499

\end{thebibliography}


\begin{thebibliography}{67}
\expandafter\ifx\csname natexlab\endcsname\relax\def\natexlab#1{#1}\fi

\bibitem[{{Antonucci} \& {Cohen}(1983)}]{antonucci83}
{Antonucci}, R.~R.~J. \& {Cohen}, R.~D. 1983, \apj, 271, 564

\bibitem[{{Arnaud}(1996)}]{Arnaud1996}
{Arnaud}, K.~A. 1996, in ASP Conf. Ser., Vol. 101, Astronomical Data Analysis Software and Systems V, ed. G.~H. {Jacoby} \& J.~{Barnes} (San Francisco: Astron. Soc. Pac.), 17--20

\bibitem[{{Arzoumanian} {et~al.}(2014){Arzoumanian}, {Gendreau}, {Baker}, {Cazeau}, {Hestnes}, {Kellogg}, {Kenyon}, {Kozon}, {Liu}, {Manthripragada}, {Markwardt}, {Mitchell}, {Mitchell}, {Monroe}, {Okajima}, {Pollard}, {Powers}, {Savadkin}, {Winternitz}, {Chen}, {Wright}, {Foster}, {Prigozhin}, {Remillard}, \& {Doty}}]{Arzoumanian14}
{Arzoumanian}, Z., {Gendreau}, K.~C., {Baker}, C.~L., {et~al.} 2014, in Society of Photo-Optical Instrumentation Engineers (SPIE) Conference Series, Vol. 9144, Space Telescopes and Instrumentation 2014: Ultraviolet to Gamma Ray, ed. T.~{Takahashi}, J.-W.~A. {den Herder}, \& M.~{Bautz}, 914420

\bibitem[{{Baldini} {et~al.}(2022){Baldini}, {Bucciantini}, {Lalla}, {Ehlert}, {Manfreda}, {Negro}, {Omodei}, {Pesce-Rollins}, {Sgr{\`o}}, \& {Silvestri}}]{2022SoftX..1901194B}
{Baldini}, L., {Bucciantini}, N., {Lalla}, N.~D., {et~al.} 2022, SoftwareX, 19, 101194

\bibitem[{{Balokovi{\'c}} {et~al.}(2018){Balokovi{\'c}}, {Brightman}, {Harrison}, {Comastri}, {Ricci}, {Buchner}, {Gandhi}, {Farrah}, \& {Stern}}]{balokovic18}
{Balokovi{\'c}}, M., {Brightman}, M., {Harrison}, F.~A., {et~al.} 2018, \apj, 854, 42

\bibitem[{{Balokovi{\'c}} {et~al.}(2019){Balokovi{\'c}}, {Garc{\'\i}a}, \& {Cabral}}]{balokovic19}
{Balokovi{\'c}}, M., {Garc{\'\i}a}, J.~A., \& {Cabral}, S.~E. 2019, Research Notes of the American Astronomical Society, 3, 173

\bibitem[{{Bentz} {et~al.}(2022){Bentz}, {Williams}, \& {Treu}}]{Bentz2022}
{Bentz}, M.~C., {Williams}, P.~R., \& {Treu}, T. 2022, \apj, 934, 168

\bibitem[{{Beuchert} {et~al.}(2017){Beuchert}, {Markowitz}, {Dauser}, {Garc{\'\i}a}, {Keck}, {Wilms}, {Kadler}, {Brenneman}, \& {Zdziarski}}]{Beuchert2017}
{Beuchert}, T., {Markowitz}, A.~G., {Dauser}, T., {et~al.} 2017, \aap, 603, A50

\bibitem[{{Bianchi} {et~al.}(2010){Bianchi}, {Chiaberge}, {Evans}, {Guainazzi}, {Baldi}, {Matt}, \& {Piconcelli}}]{Bianchi2010a}
{Bianchi}, S., {Chiaberge}, M., {Evans}, D.~A., {et~al.} 2010, \mnras, 405, 553

\bibitem[{{De Rosa} {et~al.}(2007){De Rosa}, {Piro}, {Perola}, {Capalbi}, {Cappi}, {Grandi}, {Maraschi}, \& {Petrucci}}]{derosa07}
{De Rosa}, A., {Piro}, L., {Perola}, G.~C., {et~al.} 2007, \aap, 463, 903

\bibitem[{{Di Marco} {et~al.}(2022){Di Marco}, {Costa}, {Muleri}, {Soffitta}, {Fabiani}, {La Monaca}, {Rankin}, {Xie}, {Bachetti}, {Baldini}, {Baumgartner}, {Bellazzini}, {Brez}, {Castellano}, {Del Monte}, {Di Lalla}, {Ferrazzoli}, {Latronico}, {Maldera}, {Manfreda}, {O'Dell}, {Perri}, {Pesce-Rollins}, {Puccetti}, {Ramsey}, {Ratheesh}, {Sgr{\`o}}, {Spandre}, {Tennant}, {Tobia}, {Trois}, \& {Weisskopf}}]{dimarco22}
{Di Marco}, A., {Costa}, E., {Muleri}, F., {et~al.} 2022, \aj, 163, 170

\bibitem[{{Di Marco} {et~al.}(2023){Di Marco}, {Soffitta}, {Costa}, {Ferrazzoli}, {La Monaca}, {Rankin}, {Ratheesh}, {Xie}, {Baldini}, {Del Monte}, {Ehlert}, {Fabiani}, {Kim}, {Muleri}, {O'Dell}, {Ramsey}, {Rubini}, {Sgr{\`o}}, {Silvestri}, {Tennant}, \& {Weisskopf}}]{dimarco23}
{Di Marco}, A., {Soffitta}, P., {Costa}, E., {et~al.} 2023, \aj, 165, 143

\bibitem[{{Draper} {et~al.}(1992){Draper}, {Gledhill}, {Scarrott}, \& {Tadhunter}}]{Draper1992}
{Draper}, P.~W., {Gledhill}, T.~M., {Scarrott}, S.~M., \& {Tadhunter}, C.~N. 1992, \mnras, 257, 309

\bibitem[{{Fabian} {et~al.}(2017){Fabian}, {Lohfink}, {Belmont}, {Malzac}, \& {Coppi}}]{fabian17}
{Fabian}, A.~C., {Lohfink}, A., {Belmont}, R., {Malzac}, J., \& {Coppi}, P. 2017, \mnras, 467, 2566

\bibitem[{{Fabian} {et~al.}(2015){Fabian}, {Lohfink}, {Kara}, {Parker}, {Vasudevan}, \& {Reynolds}}]{fabian15}
{Fabian}, A.~C., {Lohfink}, A., {Kara}, E., {et~al.} 2015, \mnras, 451, 4375

\bibitem[{{Ferland} {et~al.}(1998){Ferland}, {Korista}, {Verner}, {Ferguson}, {Kingdon}, \& {Verner}}]{ferland98}
{Ferland}, G.~J., {Korista}, K.~T., {Verner}, D.~A., {et~al.} 1998, \pasp, 110, 761

\bibitem[{{Gendreau} {et~al.}(2016){Gendreau}, {Arzoumanian}, {Adkins}, {Albert}, {Anders}, {Aylward}, {Baker}, {Balsamo}, {Bamford}, {Benegalrao}, {Berry}, {Bhalwani}, {Black}, {Blaurock}, {Bronke}, {Brown}, {Budinoff}, {Cantwell}, {Cazeau}, {Chen}, {Clement}, {Colangelo}, {Coleman}, {Coopersmith}, {Dehaven}, {Doty}, {Egan}, {Enoto}, {Fan}, {Ferro}, {Foster}, {Galassi}, {Gallo}, {Green}, {Grosh}, {Ha}, {Hasouneh}, {Heefner}, {Hestnes}, {Hoge}, {Jacobs}, {J{\o}rgensen}, {Kaiser}, {Kellogg}, {Kenyon}, {Koenecke}, {Kozon}, {LaMarr}, {Lambertson}, {Larson}, {Lentine}, {Lewis}, {Lilly}, {Liu}, {Malonis}, {Manthripragada}, {Markwardt}, {Matonak}, {Mcginnis}, {Miller}, {Mitchell}, {Mitchell}, {Mohammed}, {Monroe}, {Montt de Garcia}, {Mul{\'e}}, {Nagao}, {Ngo}, {Norris}, {Norwood}, {Novotka}, {Okajima}, {Olsen}, {Onyeachu}, {Orosco}, {Peterson}, {Pevear}, {Pham}, {Pollard}, {Pope}, {Powers}, {Powers}, {Price}, {Prigozhin}, {Ramirez}, {Reid}, {Remillard}, {Rogstad}, {Rosecrans}, {Rowe}, {Sager}, {Sanders},
  {Savadkin}, {Saylor}, {Schaeffer}, {Schweiss}, {Semper}, {Serlemitsos}, {Shackelford}, {Soong}, {Struebel}, {Vezie}, {Villasenor}, {Winternitz}, {Wofford}, {Wright}, {Yang}, \& {Yu}}]{gendreau16}
{Gendreau}, K.~C., {Arzoumanian}, Z., {Adkins}, P.~W., {et~al.} 2016, in Society of Photo-Optical Instrumentation Engineers (SPIE) Conference Series, Vol. 9905, Space Telescopes and Instrumentation 2016: Ultraviolet to Gamma Ray, ed. J.-W.~A. {den Herder}, T.~{Takahashi}, \& M.~{Bautz}, 99051H

\bibitem[{{Gendreau} {et~al.}(2012){Gendreau}, {Arzoumanian}, \& {Okajima}}]{Gendreau12}
{Gendreau}, K.~C., {Arzoumanian}, Z., \& {Okajima}, T. 2012, in Society of Photo-Optical Instrumentation Engineers (SPIE) Conference Series, Vol. 8443, Space Telescopes and Instrumentation 2012: Ultraviolet to Gamma Ray, ed. T.~{Takahashi}, S.~S. {Murray}, \& J.-W.~A. {den Herder}, 844313

\bibitem[{{Gianolli} {et~al.}(2024){Gianolli}, {Bianchi}, {Petrucci}, {Marinucci}, {Ingram}, {Tagliacozzo}, {Kim}, {Marin}, {Matt}, {Soffitta}, \& {Tombesi}}]{gianolli24}
{Gianolli}, V.~E., {Bianchi}, S., {Petrucci}, P.-O., {et~al.} 2024, in Memorie della Societa Astronomica Italiana, Vol.~95, 27--31

\bibitem[{{Gianolli} {et~al.}(2023){Gianolli}, {Kim}, {Bianchi}, {Ag{\'\i}s-Gonz{\'a}lez}, {Madejski}, {Marin}, {Marinucci}, {Matt}, {Middei}, {Petrucci}, {Soffitta}, {Tagliacozzo}, {Tombesi}, {Ursini}, {Barnouin}, {De Rosa}, {Di Gesu}, {Ingram}, {Loktev}, {Panagiotou}, {Podgorny}, {Poutanen}, {Puccetti}, {Ratheesh}, {Veledina}, {Zhang}, {Agudo}, {Antonelli}, {Bachetti}, {Baldini}, {Baumgartner}, {Bellazzini}, {Bongiorno}, {Bonino}, {Brez}, {Bucciantini}, {Capitanio}, {Castellano}, {Cavazzuti}, {Chen}, {Ciprini}, {Costa}, {Del Monte}, {Di Lalla}, {Di Marco}, {Donnarumma}, {Doroshenko}, {Dov{\v{c}}iak}, {Ehlert}, {Enoto}, {Evangelista}, {Fabiani}, {Ferrazzoli}, {Garc{\'\i}a}, {Gunji}, {Heyl}, {Iwakiri}, {Jorstad}, {Kaaret}, {Karas}, {Kislat}, {Kitaguchi}, {Kolodziejczak}, {Krawczynski}, {La Monaca}, {Latronico}, {Liodakis}, {Maldera}, {Manfreda}, {Marscher}, {Marshall}, {Massaro}, {Mitsuishi}, {Mizuno}, {Muleri}, {Negro}, {Ng}, {O'Dell}, {Omodei}, {Oppedisano}, {Papitto}, {Pavlov}, {Peirson}, {Perri},
  {Pesce-Rollins}, {Pilia}, {Possenti}, {Ramsey}, {Rankin}, {Roberts}, {Romani}, {Sgr{\`o}}, {Slane}, {Spandre}, {Swartz}, {Tamagawa}, {Tavecchio}, {Taverna}, {Tawara}, {Tennant}, {Thomas}, {Trois}, {Tsygankov}, {Turolla}, {Vink}, {Weisskopf}, {Wu}, {Xie}, \& {Zane}}]{gianolli23}
{Gianolli}, V.~E., {Kim}, D.~E., {Bianchi}, S., {et~al.} 2023, \mnras, 523, 4468

\bibitem[{{Harrison} {et~al.}(1986){Harrison}, {Pedlar}, {Unger}, {Burgess}, {Graham}, \& {Preuss}}]{Harrison1986}
{Harrison}, B., {Pedlar}, A., {Unger}, S.~W., {et~al.} 1986, \mnras, 218, 775

\bibitem[{{Harrison} {et~al.}(2013){Harrison}, {Craig}, {Christensen}, {Hailey}, {Zhang}, {Boggs}, {Stern}, {Cook}, {Forster}, {Giommi}, {Grefenstette}, {Kim}, {Kitaguchi}, {Koglin}, {Madsen}, {Mao}, {Miyasaka}, {Mori}, {Perri}, {Pivovaroff}, {Puccetti}, {Rana}, {Westergaard}, {Willis}, {Zoglauer}, {An}, {Bachetti}, {Barri{\`e}re}, {Bellm}, {Bhalerao}, {Brejnholt}, {Fuerst}, {Liebe}, {Markwardt}, {Nynka}, {Vogel}, {Walton}, {Wik}, {Alexander}, {Cominsky}, {Hornschemeier}, {Hornstrup}, {Kaspi}, {Madejski}, {Matt}, {Molendi}, {Smith}, {Tomsick}, {Ajello}, {Ballantyne}, {Balokovi{\'c}}, {Barret}, {Bauer}, {Blandford}, {Brandt}, {Brenneman}, {Chiang}, {Chakrabarty}, {Chenevez}, {Comastri}, {Dufour}, {Elvis}, {Fabian}, {Farrah}, {Fryer}, {Gotthelf}, {Grindlay}, {Helfand}, {Krivonos}, {Meier}, {Miller}, {Natalucci}, {Ogle}, {Ofek}, {Ptak}, {Reynolds}, {Rigby}, {Tagliaferri}, {Thorsett}, {Treister}, \& {Urry}}]{harrison13}
{Harrison}, F.~A., {Craig}, W.~W., {Christensen}, F.~E., {et~al.} 2013, \apj, 770, 103

\bibitem[{{Ingram} {et~al.}(2023){Ingram}, {Ewing}, {Marinucci}, {Tagliacozzo}, {Rosario}, {Veledina}, {Kim}, {Marin}, {Bianchi}, {Poutanen}, {Matt}, {Marshall}, {Ursini}, {De Rosa}, {Petrucci}, {Madejski}, {Barnouin}, {Gesu}, {Dov{\v{c}}iak}, {Gianolli}, {Krawczynski}, {Loktev}, {Middei}, {Podgorny}, {Puccetti}, {Ratheesh}, {Soffitta}, {Tombesi}, {Ehlert}, {Massaro}, {Agudo}, {Antonelli}, {Bachetti}, {Baldini}, {Baumgartner}, {Bellazzini}, {Bongiorno}, {Bonino}, {Brez}, {Bucciantini}, {Capitanio}, {Castellano}, {Cavazzuti}, {Chen}, {Ciprini}, {Costa}, {Del Monte}, {Lalla}, {Marco}, {Donnarumma}, {Doroshenko}, {Enoto}, {Evangelista}, {Fabiani}, {Ferrazzoli}, {Garc{\'\i}a}, {Gunji}, {Heyl}, {Iwakiri}, {Jorstad}, {Kaaret}, {Karas}, {Kislat}, {Kitaguchi}, {Kolodziejczak}, {Monaca}, {Latronico}, {Liodakis}, {Maldera}, {Manfreda}, {Marscher}, {Mitsuishi}, {Mizuno}, {Muleri}, {Negro}, {Ng}, {O'Dell}, {Omodei}, {Oppedisano}, {Papitto}, {Pavlov}, {Peirson}, {Perri}, {Pesce-Rollins}, {Pilia}, {Possenti}, {Ramsey},
  {Rankin}, {Roberts}, {Romani}, {Sgr{\`o}}, {Slane}, {Spandre}, {Swartz}, {Tamagawa}, {Tavecchio}, {Taverna}, {Tawara}, {Tennant}, {Thomas}, {Trois}, {Tsygankov}, {Turolla}, {Vink}, {Weisskopf}, {Wu}, {Xie}, \& {Zane}}]{ingram23}
{Ingram}, A., {Ewing}, M., {Marinucci}, A., {et~al.} 2023, \mnras, 525, 5437

\bibitem[{{Kaastra} \& {Bleeker}(2016)}]{kaastra16}
{Kaastra}, J.~S. \& {Bleeker}, J.~A.~M. 2016, \aap, 587, A151

\bibitem[{{Kamraj} {et~al.}(2022){Kamraj}, {Brightman}, {Harrison}, {Stern}, {Garc{\'\i}a}, {Balokovi{\'c}}, {Ricci}, {Koss}, {Mej{\'\i}a-Restrepo}, {Oh}, {Powell}, \& {Urry}}]{kamraj22}
{Kamraj}, N., {Brightman}, M., {Harrison}, F.~A., {et~al.} 2022, \apj, 927, 42

\bibitem[{{Kang} \& {Wang}(2022)}]{kang22}
{Kang}, J.-L. \& {Wang}, J.-X. 2022, \apj, 929, 141

\bibitem[{{Keck} {et~al.}(2015){Keck}, {Brenneman}, {Ballantyne}, {Bauer}, {Boggs}, {Christensen}, {Craig}, {Dauser}, {Elvis}, {Fabian}, {Fuerst}, {Garc{\'\i}a}, {Grefenstette}, {Hailey}, {Harrison}, {Madejski}, {Marinucci}, {Matt}, {Reynolds}, {Stern}, {Walton}, \& {Zoghbi}}]{keck15}
{Keck}, M.~L., {Brenneman}, L.~W., {Ballantyne}, D.~R., {et~al.} 2015, \apj, 806, 149

\bibitem[{{Kraemer} {et~al.}(2008){Kraemer}, {Schmitt}, \& {Crenshaw}}]{kraemer08}
{Kraemer}, S.~B., {Schmitt}, H.~R., \& {Crenshaw}, D.~M. 2008, \apj, 679, 1128

\bibitem[{{Lubi{\'n}ski} {et~al.}(2010){Lubi{\'n}ski}, {Zdziarski}, {Walter}, {Paltani}, {Beckmann}, {Soldi}, {Ferrigno}, \& {Courvoisier}}]{lubinski10}
{Lubi{\'n}ski}, P., {Zdziarski}, A.~A., {Walter}, R., {et~al.} 2010, \mnras, 408, 1851

\bibitem[{Marin {et~al.}(2024)Marin, Gianolli, Ingram, Kim, Marinucci, Tagliacozzo, \& Ursini}]{marin2024}
Marin, F., Gianolli, V.~E., Ingram, A., {et~al.} 2024, Galaxies, 12

\bibitem[{{Marin} {et~al.}(2020){Marin}, {Le Cam}, {Lopez-Rodriguez}, {Kolehmainen}, {Babler}, \& {Meade}}]{Marin2020}
{Marin}, F., {Le Cam}, J., {Lopez-Rodriguez}, E., {et~al.} 2020, \mnras, 496, 215

\bibitem[{{Marinucci} {et~al.}(2022){Marinucci}, {Muleri}, {Dovciak}, {Bianchi}, {Marin}, {Matt}, {Ursini}, {Middei}, {Marshall}, {Baldini}, {Barnouin}, {Rodriguez}, {De Rosa}, {Di Gesu}, {Harper}, {Ingram}, {Karas}, {Krawczynski}, {Madejski}, {Panagiotou}, {Petrucci}, {Podgorny}, {Puccetti}, {Tombesi}, {Veledina}, {Zhang}, {Agudo}, {Antonelli}, {Bachetti}, {Baumgartner}, {Bellazzini}, {Bongiorno}, {Bonino}, {Brez}, {Bucciantini}, {Capitanio}, {Castellano}, {Cavazzuti}, {Ciprini}, {Costa}, {Del Monte}, {Di Lalla}, {Di Marco}, {Donnarumma}, {Doroshenko}, {Ehlert}, {Enoto}, {Evangelista}, {Fabiani}, {Ferrazzoli}, {Garcia}, {Gunji}, {Hayashida}, {Heyl}, {Iwakiri}, {Jorstad}, {Kitaguchi}, {Kolodziejczak}, {La Monaca}, {Latronico}, {Liodakis}, {Maldera}, {Manfreda}, {Marscher}, {Mitsuishi}, {Mizuno}, {Ng}, {O'Dell}, {Omodei}, {Oppedisano}, {Papitto}, {Pavlov}, {Peirson}, {Perri}, {Pesce-Rollins}, {Pilia}, {Possenti}, {Poutanen}, {Ramsey}, {Rankin}, {Ratheesh}, {Romani}, {Sgr{\v{s}}}, {Slane}, {Soffitta},
  {Spandre}, {Tamagawa}, {Tavecchio}, {Taverna}, {Tawara}, {Tennant}, {Thomas}, {Trois}, {Tsygankov}, {Turolla}, {Vink}, {Weisskopf}, {Wu}, {Xie}, \& {Zane}}]{marinucci22}
{Marinucci}, A., {Muleri}, F., {Dovciak}, M., {et~al.} 2022, \mnras, 516, 5907

\bibitem[{{Matt} {et~al.}(2015){Matt}, {Balokovi{\'c}}, {Marinucci}, {Ballantyne}, {Boggs}, {Christensen}, {Comastri}, {Craig}, {Gandhi}, {Hailey}, {Harrison}, {Madejski}, {Madsen}, {Stern}, \& {Zhang}}]{matt15}
{Matt}, G., {Balokovi{\'c}}, M., {Marinucci}, A., {et~al.} 2015, \mnras, 447, 3029

\bibitem[{{May} {et~al.}(2020){May}, {Steiner}, {Menezes}, {Williams}, \& {Wang}}]{May2020}
{May}, D., {Steiner}, J.~E., {Menezes}, R.~B., {Williams}, D.~R.~A., \& {Wang}, J. 2020, \mnras, 496, 1488

\bibitem[{Muleri(2022)}]{Muleri2022}
Muleri, F. 2022, Analysis of the Data from Photoelectric Gas Polarimeters, ed. C.~Bambi \& A.~Santangelo (Singapore: Springer Nature Singapore), 1--23

\bibitem[{{Mundell} {et~al.}(2003){Mundell}, {Wrobel}, {Pedlar}, \& {Gallimore}}]{Mundell2003}
{Mundell}, C.~G., {Wrobel}, J.~M., {Pedlar}, A., \& {Gallimore}, J.~F. 2003, \apj, 583, 192

\bibitem[{{NASA Heasarc}(2014)}]{2014ascl.soft08004N}
{NASA Heasarc}. 2014, {HEAsoft: Unified Release of FTOOLS and XANADU}, Astrophysics Source Code Library, record ascl:1408.004

\bibitem[{{Petrucci} {et~al.}(2001){Petrucci}, {Haardt}, {Maraschi}, {Grandi}, {Malzac}, {Matt}, {Nicastro}, {Piro}, {Perola}, \& {De Rosa}}]{petrucci01}
{Petrucci}, P.~O., {Haardt}, F., {Maraschi}, L., {et~al.} 2001, \apj, 556, 716

\bibitem[{{Piconcelli} {et~al.}(2004){Piconcelli}, {Jimenez-Bail{\'o}n}, {Guainazzi}, {Schartel}, {Rodr{\'\i}guez-Pascual}, \& {Santos-Lle{\'o}}}]{piconcelli04}
{Piconcelli}, E., {Jimenez-Bail{\'o}n}, E., {Guainazzi}, M., {et~al.} 2004, \mnras, 351, 161

\bibitem[{{Podgorn{\'y}} {et~al.}(2023){Podgorn{\'y}}, {Dov{\v{c}}iak}, {Goosmann}, {Marin}, {Matt}, {R{\'o}{\.z}a{\'n}ska}, \& {Karas}}]{Podgorny2023}
{Podgorn{\'y}}, J., {Dov{\v{c}}iak}, M., {Goosmann}, R., {et~al.} 2023, \mnras, 524, 3853

\bibitem[{{Podgorn{\'y}} {et~al.}(2022){Podgorn{\'y}}, {Dov{\v{c}}iak}, {Marin}, {Goosmann}, \& {R{\'o}{\.z}a{\'n}ska}}]{Podgorny2022}
{Podgorn{\'y}}, J., {Dov{\v{c}}iak}, M., {Marin}, F., {Goosmann}, R., \& {R{\'o}{\.z}a{\'n}ska}, A. 2022, \mnras, 510, 4723

\bibitem[{{Ricci} {et~al.}(2018){Ricci}, {Ho}, {Fabian}, {Trakhtenbrot}, {Koss}, {Ueda}, {Lohfink}, {Shimizu}, {Bauer}, {Mushotzky}, {Schawinski}, {Paltani}, {Lamperti}, {Treister}, \& {Oh}}]{ricci18}
{Ricci}, C., {Ho}, L.~C., {Fabian}, A.~C., {et~al.} 2018, \mnras, 480, 1819

\bibitem[{{Schnittman} \& {Krolik}(2010)}]{Schnittman10}
{Schnittman}, J.~D. \& {Krolik}, J.~H. 2010, \apj, 712, 908

\bibitem[{{Serafinelli} {et~al.}(2024){Serafinelli}, {De Rosa}, {Tortosa}, {Stella}, {Vagnetti}, {Bianchi}, {Ricci}, {Kammoun}, {Petrucci}, {Middei}, {Lanzuisi}, {Marinucci}, {Ursini}, \& {Matt}}]{serafinelli24}
{Serafinelli}, R., {De Rosa}, A., {Tortosa}, A., {et~al.} 2024, arXiv e-prints, arXiv:2407.06769

\bibitem[{{Serafinelli} {et~al.}(2023){Serafinelli}, {Marinucci}, {De Rosa}, {Bianchi}, {Middei}, {Matt}, {Reeves}, {Braito}, {Tombesi}, {Gianolli}, {Ingram}, {Marin}, {Petrucci}, {Tagliacozzo}, \& {Ursini}}]{serafinelli23}
{Serafinelli}, R., {Marinucci}, A., {De Rosa}, A., {et~al.} 2023, \mnras, 526, 3540

\bibitem[{{Shapiro} {et~al.}(1976){Shapiro}, {Lightman}, \& {Eardley}}]{shapiro76}
{Shapiro}, S.~L., {Lightman}, A.~P., \& {Eardley}, D.~M. 1976, \apj, 204, 187

\bibitem[{{Shapovalova} {et~al.}(2012){Shapovalova}, {Popovic}, {Collin}, {Burenkov}, {Chavushyan}, {Bochkarev}, {Ilic}, {Kovacevic}, \& {Mercado}}]{shapovalova12}
{Shapovalova}, A.~I., {Popovic}, L.~C., {Collin}, S., {et~al.} 2012, Astron. Astrophys. Trans., 27, 429

\bibitem[{{Storchi-Bergmann} {et~al.}(2009){Storchi-Bergmann}, {McGregor}, {Riffel}, {Sim{\~o}es Lopes}, {Beck}, \& {Dopita}}]{Storchi-Bergmann2009}
{Storchi-Bergmann}, T., {McGregor}, P.~J., {Riffel}, R.~A., {et~al.} 2009, \mnras, 394, 1148

\bibitem[{{Str{\"u}der} {et~al.}(2001){Str{\"u}der}, {Briel}, {Dennerl}, {Hartmann}, {Kendziorra}, {Meidinger}, {Pfeffermann}, {Reppin}, {Aschenbach}, {Bornemann}, {Br{\"a}uninger}, {Burkert}, {Elender}, {Freyberg}, {Haberl}, {Hartner}, {Heuschmann}, {Hippmann}, {Kastelic}, {Kemmer}, {Kettenring}, {Kink}, {Krause}, {M{\"u}ller}, {Oppitz}, {Pietsch}, {Popp}, {Predehl}, {Read}, {Stephan}, {St{\"o}tter}, {Tr{\"u}mper}, {Holl}, {Kemmer}, {Soltau}, {St{\"o}tter}, {Weber}, {Weichert}, {von Zanthier}, {Carathanassis}, {Lutz}, {Richter}, {Solc}, {B{\"o}ttcher}, {Kuster}, {Staubert}, {Abbey}, {Holland}, {Turner}, {Balasini}, {Bignami}, {La Palombara}, {Villa}, {Buttler}, {Gianini}, {Lain{\'e}}, {Lumb}, \& {Dhez}}]{struder01}
{Str{\"u}der}, L., {Briel}, U., {Dennerl}, K., {et~al.} 2001, \aap, 365, L18

\bibitem[{{Sunyaev} \& {Titarchuk}(1980)}]{sunyaev80}
{Sunyaev}, R.~A. \& {Titarchuk}, L.~G. 1980, \aap, 86, 121

\bibitem[{{Tagliacozzo} {et~al.}(2023){Tagliacozzo}, {Marinucci}, {Ursini}, {Matt}, {Bianchi}, {Baldini}, {Barnouin}, {Cavero Rodriguez}, {De Rosa}, {Di Gesu}, {Dov{\v{c}}iak}, {Harper}, {Ingram}, {Karas}, {Kim}, {Krawczynski}, {Madejski}, {Marin}, {Middei}, {Marshall}, {Muleri}, {Panagiotou}, {Petrucci}, {Podgorny}, {Poutanen}, {Puccetti}, {Soffitta}, {Tombesi}, {Veledina}, {Zhang}, {Agudo}, {Antonelli}, {Bachetti}, {Baumgartner}, {Bellazzini}, {Bongiorno}, {Bonino}, {Brez}, {Bucciantini}, {Capitanio}, {Castellano}, {Cavazzuti}, {Chen}, {Ciprini}, {Costa}, {Del Monte}, {Di Lalla}, {Di Marco}, {Donnarumma}, {Doroshenko}, {Ehlert}, {Enoto}, {Evangelista}, {Fabiani}, {Ferrazzoli}, {Garcia}, {Gunji}, {Heyl}, {Iwakiri}, {Jorstad}, {Kaaret}, {Kislat}, {Kitaguchi}, {Kolodziejczak}, {La Monaca}, {Latronico}, {Liodakis}, {Maldera}, {Manfreda}, {Marscher}, {Massaro}, {Mitsuishi}, {Mizuno}, {Negro}, {Ng}, {O'Dell}, {Omodei}, {Oppedisano}, {Papitto}, {Pavlov}, {Peirson}, {Perri}, {Pesce-Rollins}, {Pilia}, {Possenti},
  {Ramsey}, {Rankin}, {Ratheesh}, {Roberts}, {Romani}, {Sgr{\`o}}, {Slane}, {Spandre}, {Swartz}, {Tamagawa}, {Tavecchio}, {Taverna}, {Tawara}, {Tennant}, {Thomas}, {Trois}, {Tsygankov}, {Turolla}, {Vink}, {Weisskopf}, {Wu}, {Xie}, \& {Zane}}]{tagliacozzo23}
{Tagliacozzo}, D., {Marinucci}, A., {Ursini}, F., {et~al.} 2023, \mnras, 525, 4735

\bibitem[{{Tamborra} {et~al.}(2018){Tamborra}, {Matt}, {Bianchi}, \& {Dov{\v{c}}iak}}]{tamborra18}
{Tamborra}, F., {Matt}, G., {Bianchi}, S., \& {Dov{\v{c}}iak}, M. 2018, \aap, 619, A105

\bibitem[{{Tortosa} {et~al.}(2018){Tortosa}, {Bianchi}, {Marinucci}, {Matt}, \& {Petrucci}}]{tortosa18}
{Tortosa}, A., {Bianchi}, S., {Marinucci}, A., {Matt}, G., \& {Petrucci}, P.~O. 2018, \aap, 614, A37

\bibitem[{{Ulvestad} {et~al.}(1998){Ulvestad}, {Roy}, {Colbert}, \& {Wilson}}]{ulvestad98}
{Ulvestad}, J.~S., {Roy}, A.~L., {Colbert}, E. J.~M., \& {Wilson}, A.~S. 1998, \apj, 496, 196

\bibitem[{{Ursini} {et~al.}(2022){Ursini}, {Matt}, {Bianchi}, {Marinucci}, {Dov{\v{c}}iak}, \& {Zhang}}]{ursini22}
{Ursini}, F., {Matt}, G., {Bianchi}, S., {et~al.} 2022, \mnras, 510, 3674

\bibitem[{{Wang} {et~al.}(2011){Wang}, {Fabbiano}, {Risaliti}, {Elvis}, {Karovska}, {Zezas}, {Mundell}, {Dumas}, \& {Schinnerer}}]{Wang2011a}
{Wang}, J., {Fabbiano}, G., {Risaliti}, G., {et~al.} 2011, \apj, 729, 75

\bibitem[{{Weaver} {et~al.}(1994){Weaver}, {Mushotzky}, {Arnaud}, {Serlemitsos}, {Marshall}, {Petre}, {Jahoda}, {Smale}, \& {Netzer}}]{weaver94}
{Weaver}, K.~A., {Mushotzky}, R.~F., {Arnaud}, K.~A., {et~al.} 1994, \apj, 423, 621

\bibitem[{{Weisskopf} {et~al.}(2022){Weisskopf}, {Soffitta}, {Baldini}, {Ramsey}, {O'Dell}, {Romani}, {Matt}, {Deininger}, {Baumgartner}, {Bellazzini}, {Costa}, {Kolodziejczak}, {Latronico}, {Marshall}, {Muleri}, {Bongiorno}, {Tennant}, {Bucciantini}, {Dovciak}, {Marin}, {Marscher}, {Poutanen}, {Slane}, {Turolla}, {Kalinowski}, {Di Marco}, {Fabiani}, {Minuti}, {La Monaca}, {Pinchera}, {Rankin}, {Sgro'}, {Trois}, {Xie}, {Alexander}, {Allen}, {Amici}, {Andersen}, {Antonelli}, {Antoniak}, {Attin{\`a}}, {Barbanera}, {Bachetti}, {Baggett}, {Bladt}, {Brez}, {Bonino}, {Boree}, {Borotto}, {Breeding}, {Brienza}, {Bygott}, {Caporale}, {Cardelli}, {Carpentiero}, {Castellano}, {Castronuovo}, {Cavalli}, {Cavazzuti}, {Ceccanti}, {Centrone}, {Citraro}, {D'Amico}, {D'Alba}, {Di Gesu}, {Del Monte}, {Dietz}, {Di Lalla}, {Persio}, {Dolan}, {Donnarumma}, {Evangelista}, {Ferrant}, {Ferrazzoli}, {Ferrie}, {Footdale}, {Forsyth}, {Foster}, {Garelick}, {Gunji}, {Gurnee}, {Head}, {Hibbard}, {Johnson}, {Kelly}, {Kilaru}, {Lefevre},
  {Roy}, {Loffredo}, {Lorenzi}, {Lucchesi}, {Maddox}, {Magazzu}, {Maldera}, {Manfreda}, {Mangraviti}, {Marengo}, {Marrocchesi}, {Massaro}, {Mauger}, {McCracken}, {McEachen}, {Mize}, {Mereu}, {Mitchell}, {Mitsuishi}, {Morbidini}, {Mosti}, {Nasimi}, {Negri}, {Negro}, {Nguyen}, {Nitschke}, {Nuti}, {Onizuka}, {Oppedisano}, {Orsini}, {Osborne}, {Pacheco}, {Paggi}, {Painter}, {Pavelitz}, {Pentz}, {Piazzolla}, {Perri}, {Pesce-Rollins}, {Peterson}, {Pilia}, {Profeti}, {Puccetti}, {Ranganathan}, {Ratheesh}, {Reedy}, {Root}, {Rubini}, {Ruswick}, {Sanchez}, {Sarra}, {Santoli}, {Scalise}, {Sciortino}, {Schroeder}, {Seek}, {Sosdian}, {Spandre}, {Speegle}, {Tamagawa}, {Tardiola}, {Tobia}, {Thomas}, {Valerie}, {Vimercati}, {Walden}, {Weddendorf}, {Wedmore}, {Welch}, {Zanetti}, \& {Zanetti}}]{weisskopf22}
{Weisskopf}, M.~C., {Soffitta}, P., {Baldini}, L., {et~al.} 2022, J. Astron. Telesc. Instrum. Syst., 8, 026002

\bibitem[{{Williams} {et~al.}(2017){Williams}, {McHardy}, {Baldi}, {Beswick}, {Argo}, {Dullo}, {Knapen}, {Brinks}, {Fenech}, {Mundell}, {Muxlow}, {Panessa}, {Rampadarath}, \& {Westcott}}]{Williams2017}
{Williams}, D.~R.~A., {McHardy}, I.~M., {Baldi}, R.~D., {et~al.} 2017, \mnras, 472, 3842

\bibitem[{{Wolfinger} {et~al.}(2013){Wolfinger}, {Kilborn}, {Koribalski}, {Minchin}, {Boyce}, {Disney}, {Lang}, \& {Jordan}}]{wolfinger13}
{Wolfinger}, K., {Kilborn}, V.~A., {Koribalski}, B.~S., {et~al.} 2013, \mnras, 428, 1790

\bibitem[{{Yang} {et~al.}(2001){Yang}, {Wilson}, \& {Ferruit}}]{yang01}
{Yang}, Y., {Wilson}, A.~S., \& {Ferruit}, P. 2001, \apj, 563, 124

\bibitem[{{Yaqoob} {et~al.}(1995){Yaqoob}, {Edelson}, {Weaver}, {Warwick}, {Mushotzky}, {Serlemitsos}, \& {Holt}}]{yaqoob95}
{Yaqoob}, T., {Edelson}, R., {Weaver}, K.~A., {et~al.} 1995, \apjl, 453, L81

\bibitem[{{Zdziarski} {et~al.}(1996){Zdziarski}, {Johnson}, \& {Magdziarz}}]{zdziarski96}
{Zdziarski}, A.~A., {Johnson}, W.~N., \& {Magdziarz}, P. 1996, \mnras, 283, 193

\bibitem[{{Zdziarski} {et~al.}(2000){Zdziarski}, {Poutanen}, \& {Johnson}}]{zdziarski00}
{Zdziarski}, A.~A., {Poutanen}, J., \& {Johnson}, W.~N. 2000, \apj, 542, 703

\bibitem[{{Zhang} {et~al.}(2019){Zhang}, {Dov{\v{c}}iak}, \& {Bursa}}]{zhang19}
{Zhang}, W., {Dov{\v{c}}iak}, M., \& {Bursa}, M. 2019, \apj, 875, 148

\bibitem[{{Zoghbi} {et~al.}(2019){Zoghbi}, {Miller}, \& {Cackett}}]{zoghbi19}
{Zoghbi}, A., {Miller}, J.~M., \& {Cackett}, E. 2019, \apj, 884, 26

\bibitem[{{{\.Z}ycki} {et~al.}(1999){{\.Z}ycki}, {Done}, \& {Smith}}]{zycki99}
{{\.Z}ycki}, P.~T., {Done}, C., \& {Smith}, D.~A. 1999, \mnras, 309, 561

\end{thebibliography}
%

\noindent \textbf{Affiliations:} \\
\noindent
     $^{1}$Université Grenoble Alpes, CNRS, IPAG, 38000 Grenoble, France \\
     $^{2}$Dipartimento di Matematica e Fisica, Università degli Studi Roma Tre, Via della Vasca Navale 84, 00146 Roma, Italy \\
     $^{3}$ASI - Agenzia Spaziale Italiana, Via del Politecnico snc, 00133 Roma, Italy \\
     $^{4}$INAF – Astronomical Observatory of Rome, Via Frascati 33, 00040 Monte Porzio Catone, Italy \\
     $^{5}$INAF Istituto di Astrofisica e Planetologia Spaziali, Via del Fosso del Cavaliere 100, I-00133 Roma, Italy\\
     $^{6}$Dipartimento di Fisica, Universit\`a degli Studi di Roma “La Sapienza”, Piazzale Aldo Moro 5, I-00185 Roma, Italy\\
     $^{7}$Dipartimento di Fisica, Università degli Studi di Roma “Tor Vergata”, Via della Ricerca Scientifica 1, I-00133 Roma, Italy\\
     $^{8}$Université de Strasbourg, CNRS, Observatoire Astronomique de Strasbourg, UMR 7550, 67000 Strasbourg, France \\
     $^{9}$Centre for Extragalactic Astronomy, Department of Physics, University of Durham, South Road, Durham DH1 3LE, UK \\
     $^{10}$Institute of Space and Astronautical Science (ISAS), Japan Aerospace Exploration Agency (JAXA), Kanagawa 252-5210, Japan \\
     $^{11}$MIT Kavli Institute for Astrophysics and Space Research, Massachusetts Institute of Technology, 77 Massachusetts Avenue, Cambridge, MA 02139, USA \\
     $^{12}$Space Science Data Center, Agenzia Spaziale Italiana, Via del Politecnico snc, 00133 Roma, Italy \\
     $^{13}$Astronomical Institute of the Czech Academy of Sciences, Bocˇní II 1401/1, 14100 Praha 4, Czech Republic \\
     $^{14}$Istituto Nazionale di Fisica Nucleare, Sezione di Roma "Tor Vergata", Via della Ricerca Scientifica 1, 00133 Roma, Italy \\
     $^{15}$Department of Astronomy, University of Maryland, College Park, Maryland 20742, USA 

\end{document}